\documentclass[manuscript, review=false, screen]{acmart}

\def\BibTeX{{\rm B\kern-.05em{\sc i\kern-.025em b}\kern-.08emT\kern-.1667em\lower.7ex\hbox{E}\kern-.125emX}}

\usepackage{booktabs} 

\usepackage[ruled]{algorithm2e} 

\SetAlFnt{\small}
\SetAlCapFnt{\small}
\SetAlCapNameFnt{\small}
\SetAlCapHSkip{0pt}
\IncMargin{-\parindent}

\acmJournal{TAAS}
\acmVolume{9}
\acmNumber{4}
\acmArticle{39}
\acmYear{2010}
\acmMonth{3}
\copyrightyear{2009}

\setcopyright{none}

\acmDOI{0000001.0000001}

\received{September 2018}

\usepackage[utf8]{inputenc}

\usepackage[T1]{fontenc}

\usepackage{amssymb}
\usepackage{amsmath}

\usepackage{caption}
\usepackage{subcaption}
\usepackage[autostyle]{csquotes}
\usepackage{color}
\usepackage{todonotes}

\usepackage{tikz}
\usepackage{tkz-graph}
\usetikzlibrary{shapes.geometric, arrows, arrows.meta}
\usetikzlibrary{positioning}
\tikzstyle{process} = [rectangle, rounded corners, minimum width=3cm, minimum height=1cm,text centered, draw=black, fill=cyan!30]
\tikzstyle{arrow} = [thick,->,>=stealth]

\DeclareMathOperator*{\ld}{ld}

\usepackage[]{algorithm2e}

\usepackage{afterpage}

\begin{document}
\title{On the Detection of Mutual Influences and Their Consideration in Reinforcement Learning Processes}

\author{Stefan Rudolph}
\affiliation{%
  \institution{University of Augsburg}
  \streetaddress{Eichleitnerstr. 30}
  \city{Augsburg}
  \postcode{86159}
  \country{Germany}}
\email{stefan.rudolph@informatik.uni-augsburg.de}
\author{Sven Tomforde}
\affiliation{%
  \institution{University of Kassel}
  \city{Kassel}
  \country{Germany}
}
\email{stomforde@uni-kassel.de}
\author{Jörg Hähner}
\affiliation{%
 \institution{University of Augsburg}
  \city{Augsburg}
\country{Germany}
}
\email{joerg.haehner@informatik.uni-augsburg.de}

\begin{abstract}
Self-adaptation has been proposed as a mechanism to counter complexity in control problems of technical systems. A major driver behind self-adaptation is the idea to transfer traditional design-time decisions to runtime and into the responsibility of systems themselves. In order to deal with unforeseen events and conditions, systems need creativity -- typically realized by means of machine learning capabilities. Such learning mechanisms are based on different sources of knowledge. Feedback from the environment used for reinforcement purposes is probably the most prominent one within the self-adapting and self-organizing (SASO) systems community. However, the impact of other (sub-)systems on the success of the individual system's learning performance has mostly been neglected in this context.

In this article, we propose a novel methodology to identify effects of actions performed by other systems in a shared environment on the utility achievement of an autonomous system. Consider smart cameras (SC) as illustrating example: For goals such as 3D reconstruction of objects, the most promising configuration of one SC in terms of pan/tilt/zoom parameters depends largely on the configuration of other SCs in the vicinity. Since such mutual influences cannot be pre-defined for dynamic systems, they have to be learned at runtime. Furthermore, they have to be taken into consideration when self-improving the own configuration decisions based on a feedback loop concept, e.g., known from the SASO domain or the Autonomic and Organic Computing initiatives.

We define a methodology to detect such influences at runtime, present an approach to consider this information in a reinforcement learning technique, and analyze the behavior in artificial as well as real-world SASO system settings. 
\end{abstract}

%
%


%
%


\maketitle

\renewcommand{\shortauthors}{S. Rudolph et al.}

%
%
\section{Introduction}\label{sec:100-introduction}

The research field of self-adaptive and self-organizing (SASO) systems emerged as response to the dramatic change Information and Communication Technology (ICT) has undergone within the last two decades: Back then we started with single isolated and fully comprehensible systems with well defined system boundaries. Afterwards, ICT components have been continuously coupled with each other and single monolithic solutions have been replaced by collections of autonomous entities or \textit{agents} that communicate with each other and cooperatively solve tasks. Driven by initiatives such as \textit{Autonomic Computing}~\cite{KephartC2003a}, \textit{Organic Computing}~\cite{Mueller-SchloerT2018}, \textit{Proactive Computing}~\cite{Tennenhouse2000}, \textit{Multi-Agent Systems}~\cite{Wooldridge2009}, or \textit{Collective Adaptive Systems}~\cite{KernbachST2011} a shift in responsibilities has been established: from design-time to runtime and from engineers to systems themselves. As a result, various architectures and techniques have been established that allow for runtime self-adaptation in technical systems~\cite{Weyns2013}.

These autonomous systems typically act in shared environments with the goal to achieve and maintain a given utility function by means of adapting their configuration or parameterization in response to internal and external changes. This is often combined with learning mechanisms to improve the adaptation behavior over time, which also allows for handling unexpected or disturbed conditions \cite{TomfordeM2014,Krupitzer2015}. However, these existing techniques mostly neglect that configuration and behavior of other systems in the environment may have strong impact on the degree of goal or utility achievement. These impacts become even more challenging when considering system-of-system compositions~\cite{Maier1998} or Interwoven System constellations~\cite{TomfordeHSRSWS2014} where dependencies between systems emerge that affect different abstraction layers or application domains. An illustrating example for the latter case is the coupling of ICT with the smart grid~\cite{TomfordeRBW2016}.

Consider a surveillance network consisting of a potentially large number of smart cameras (SCs) \cite{RudolphETH2014,Piciarelli2016} as illustrating example: Each SC acts autonomously and tries to optimize, e.g., the coverage of the observed area, the tracking success of conspicuous objects, or the quality of 3D reconstruction of objects. In all three goal functions, there is an impact of the current configuration of other SCs in the vicinity: The field of view should either have no overlap with the own (coverage), should have overlap at a specific position and time (tracking), or full overlap as result of a certain ideal configuration (3D reconstruction).

What we can see in this simple example is that the success of the self-adaptation strategy has not just a local origin that is under control of the system itself---it is highly influenced by the current configuration of other systems. We call these effects ``mutual influences'' to highlight that these influences stem from other systems and not the environment or possibly other sources. The shift of decisions from design-time to runtime also implies that such mutual influences cannot be fully foreseen and planned anymore. Consequently, systems need a methodology to detect them in the first place. This also requires mechanisms to incorporate this information in the self-adaptation strategy and to consider this in the self-learning behavior.

In order to be applicable to SASO systems in general, such a mutual influence methodology has to fulfill several requirements that are outlined in the following:
\begin{itemize}
	\item \textbf{Runtime capability}: Due to the complexity that is present in modern systems, it is a more and more challenging task to foresee all situations a system will face during its life time. It becomes virtually impossible because of the appearance and vanishing of other systems that affect the outcome of certain processes. Therefore, an influence detection algorithm has to be executable during the runtime of a system to allow for an adaption to unanticipated interaction partner, such as new devices from other stakeholders, manufacturers, or owners.
	
	\item \textbf{Heterogeneity}: To reach the goal of self-integration~\cite{Bellman2018}, a system faces the challenge to interact with several other systems that can be very heterogeneous in several aspects, e.g., scale, virtual/physical components, or connectedness. A special focus of this article is on the heterogeneity induced by the ownership of a system which limits the access and control of the other systems.
	
	\item \textbf{Compatibility}: Today, we face several long-living systems that are in operation for over a decade. Replacing such systems would be costly and time intensive. Therefore, to integrate with such systems, it is necessary to keep the influence detection compatible to a wide-range of systems. This especially includes systems that do not have a cooperation mechanism, are non-adaptive, or use different adaption techniques.
	
	\item \textbf{Autonomy}: Recently, we can observe a dramatic increase in interconnected ICT system surrounding us. These structures of systems bear new challenges since it is much more difficult to predict their behavior. To avoid time- and cost-intensive maintenance of these systems, we require an influence detection mechanism that can become essential part of autonomous systems, i.e., that can be integrated in a system and afterwards decide on its own which influences have to be addressed.
\end{itemize}

In this article, we present a novel methodology that fulfills these requirements. We refine our preliminary work on mutual influence detection~\cite{RudolphHTH2016,RudolphTH2016,RudolphTSH2015} and present a novel influence-aware learning mechanism. In particular, this article goes far beyond preliminary work by providing the following contributions:
\begin{itemize}
	\item A redefined method for the detection of influences among systems including multi component influences that extends previous work presented in~\cite{RudolphTSH2015} {\color{black}by introducing a general workflow and more complex use cases} (see Section~\ref{sec:300-detection-of-influences})
	\item A {novel extended method} of this approach that is runtime capable and consequently distinguishes correlations in utility and performance with other's configuration from noisy or coincidental effects (see Section~\ref{sec:300-detection-of-influences_further-aspects_runtime})
	\item A {novel approach} for incorporation of this mutual influence information within a reinforcement learning algorithm that is applied to the problem of self-adaptation of the system configuration (see Section~\ref{sec:300-detection-of-influences_further-aspects_adaption})
\end{itemize}

%
%
\section{System Model}\label{sec:200-system-model}
As basis for this article, we describe the system model and the requirements a system has to fulfill to allow for runtime influence detection. Furthermore, we introduce three example applications for the methodology that serve for illustration purposes throughout this article.

%
%
\subsection{Target Systems}\label{sec:200-system-model_target-system}
We assume an overall system that is a composition of subsystems $A_1,\dots,A_n$ in a virtual or physical environment. We refer to the term \textit{(sub)system} using the terminology from the Organic Computing domain~\cite{Mueller-SchloerT2018}, and, for a better readability, omit the \textit{sub} if it is clear from the context that not the overall system is meant. However, one might prefer to use the terms \textit{entity} or \textit{agent} instead. Each subsystem in the overall system can assume different configurations. Such a configuration typically consists of different components. We define the entire configuration space of a subsystem $A_i$ as Cartesian product $C_i=c_{i1}\times\dots\times c_{im}$, where $c_{ij}$ are the components of the configuration. 

Consider, e.g., a router $A_1$ in a computer network as an illustrating example following the ideas of \cite{THH10-a}. It can take varying configurations into account, such as the processed network protocol or parameter settings. E.g., an interval $c_{11} = [0,100]$ for the timeout parameter in seconds and the set $c_{12}=\{1,2,\dots,16\}$ for the buffer size in kilobyte. The entire configuration space of system $A_1$ would then be $C_1=[0,100]\times\{1,2,\dots,16\}$.

A further assumption is that the particular configurations of individual systems are non-overlapping, meaning each subsystem has its own set of configurations and cannot control those of other subsystems. This does not mean that the configuration components have to be completely disjoint in structure and values of the contained variables. For instance, two subsystems might have the same capabilities, which would lead to the same set of possible configurations in these attributes but for different subsystems. Such a relation is explicitly allowed within the model.

Besides the configuration space, we need to consider the local reward. In order to apply the proposed method, each subsystem has to estimate the success of its decisions at runtime---as a response to actions taken before. This is realized based on a feedback mechanism, with feedback possibly stemming from the environment of the subsystem (i.e. direct feedback) or from manual reward assignments (i.e. indirect feedback). This resembles the classic reinforcement model~\cite{SuttonB1998}, where the existence of such a reward is one of the basic assumptions. If there is no obvious way to create such a signal it can be useful to apply more structured approaches from the field of goal-oriented requirements engineering~\cite{Lamsweerde2001}, such as the \textit{knowledge acquisition in automated specification} (KAOS) goal model~\cite{Lamsweerde2009} that has been applied to similar tasks~\cite{FredericksDC2014}. Looking at the router example, a useful local reward can have different forms depending on the application scenario. For instance, a useful measure of success could be the throughput (see \cite{TH11-a}).

%
%
\textbf{Relations to Reinforcement Learning:}\label{sec:200-system-model_relations_to_RL}
{
\tikzstyle{agent} = [rectangle, rounded corners, ultra thick,  minimum width=3cm, minimum height=1cm,text centered, draw=black, fill=ACMOrange!30]
\tikzstyle{environment} = [rectangle, rounded corners, ultra thick,  minimum width=3cm, minimum height=1cm,text centered, draw=black, fill=ACMLightBlue!30]
\tikzstyle{arrowRL} = [ultra thick,->,>=stealth]

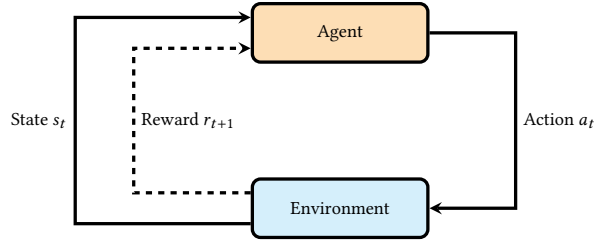
\begin{figure}[t]
    \centering
    \resizebox{8cm}{!}{%
    \begin{tikzpicture}[node distance=3cm]
        \node (agent) [agent] {Agent};
        \node (environment) [environment, below of=agent] {Environment};

        \draw [arrowRL] (agent) --++(+3,0) |- node [pos=0.25,right] {Action $a_t$} (environment);

        \path (environment.west) -- (environment.north west) coordinate[pos=0.5] (environmentHelperTop);
        \path (agent.west) -- (agent.south west) coordinate[pos=0.5] (agentHelperBottom);
        \draw [arrowRL,dashed] (environmentHelperTop) --++ (-2,0) |- node [pos=0.25,right] {Reward $r_{t+1}$} (agentHelperBottom);
        
        \path (environment.west) -- (environment.south west) coordinate[pos=0.5] (environmentHelperBottom);
        \path (agent.west) -- (agent.north west) coordinate[pos=0.5] (agentHelperTop);
        \draw [arrowRL] (environmentHelperBottom) --++ (-3,0) |- node [pos=0.25,left] {State $s_t$} (agentHelperTop);
    \end{tikzpicture}
    }
    \caption{The basic reinforcement learning model.}
    \label{fig:reinforcement-learning}
\end{figure}
}
We resemble the basic reinforcement learning~\cite{SuttonB1998}~(RL) model and its extension to multi-agent reinforcement learning~\cite{BusoniuBD2008}~(MARL). The following paragraphs discuss show how this relates to the model used in this article.

Figure~\ref{fig:reinforcement-learning} illustrates this basic model for RL. It consists of an agent that interacts with its environment. It has the ability to sense the current state of the environment and manipulate it by applying an action to it. Furthermore, a reward is provided to the agent which reflects the usefulness of the applied action, which is possibly related to the current or resulting state and can have stochastic components. The goal of the agent is to maximize the expectation of this reward over the runtime of the system by choosing the best actions available in each state. To find such a strategy, several approaches have been presented~\cite{WieringO2014}. The model is usually formalized as a Markov Decision Process (MDP), which is a tuple $(S,A,T,r)$ defining the following components:
\begin{itemize}
	\item a set of states $S$,
	\item a set of actions $A$,
	\item a transition function $T: S\times A\times S\rightarrow [0,1], T(s',a,s)\mapsto p(s'|s,a)$, i.e., a function that gives the probability that action $a$ in the state $s$ results in the state~$s'$.
	\item a reward function $r: S\times A\times S\rightarrow \mathbb{R}$, which gives a reward for each state transition.
\end{itemize}
Its basic form assumes discrete time steps $t\in\mathbb{N}$. In each time step $t$, the agent senses the environment's state $s_t\in S$. It then selects and applies an action $a_t\in A$. In the next step, $t+1$, the agent receives a reward $r_{t+1}$ which reflects the quality of its action in the given state. It then senses the new state $s_{t+1}$ and starts over again. Looking at the transition function $T$, we see that the probability of the appearance of a specific state can depend on the previous state and the selected action. The agent's goal is to find a \textit{policy} $\pi(s_t)$ that maximizes the expected discounted reward
\begin{equation}
\sum^\infty_{t=0}\gamma^tr(s_{t},a_t,s_{t+1}),
\end{equation}
where $0<\gamma<1$ is a discount factor for the future rewards.

A widely used and intensively studied RL algorithm is Q-learning 
\cite{WatkinsD1992}. Like most RL techniques, Q-learning tries to solve the general RL problem, i.e.,\ to find an optimal policy for a given problem with respect to the long term reward. The main idea is to find a Quality-function $Q:S\times A\rightarrow \mathbb{R}$ that approximates the reward for each state-action pair and takes into account the long term reward. To reach this goal, the value for each state-action-pair is initialized according to some of the various proposed methods, e.g.,\ they are all set to a fixed value or they are set to a random value, and updated afterwards according to the rule
\begin{equation}
\label{eq:q-learning}
Q_{t+1}(s_{t},a_{t})=Q_t(s_t,a_t)+\alpha\left(r_{t+1}+\gamma \max\limits_{a}Q_t(s_{t+1},a)-Q_t(s_t,a_t)\right),
\end{equation}
where $Q_t(s,a)$ denotes the old \textit{Q}-Value and $Q_{t+1}(s,a)$ the new one, each for a given state-action pair $(s,a)$. Furthermore, $r_{t+1}$ denotes the reward received in time step $t+1$ and therefore is the immediate reward for the action $a_t$ taken in time step $t$. The discount factor $\gamma\in[0,1)$ determines the fraction of estimated future rewards that is taken into account in the present step. The learning rate $\alpha\in (0,1]$ determines how much the current experience, i.e.\ the current reward, is taken into account for approximating the \textit{Q}-value. 

An extension to the single-agent RL is MARL~\cite{BusoniuBD2008}, where several agents interact with a single environment. 
Each of these agents is considered to observe and manipulate only specific parts of the environment, resulting in the agents mutually influencing their rewards. The formal approach is a generalization of the MDP and is called a stochastic game (SG). It is defined by a tuple $(S,A_1,\dots,A_n,T,r_1,\dots,r_n)$ where $n$ is the number of agents. The components are defined analogously to the single-agent case. Such SGs are classified depending on the reward structure. They are called fully cooperative if $r_1\equiv\ldots\equiv r_n$ and fully competitive if $r_1(s',a_{1},\ldots,a_{n},s)+\ldots+r_n(s',a_{1},\ldots,a_{n},s)=0$, where $a_i$ is the action of the $i$-th agent. Games between these border cases are called mixed games.

We see several similarities between the initial system model and that from the RL and MARL domains. The initially discussed configuration is similar to the action set since both are encoding the \textit{choice} of the agent/subsystem. 
However, while actions are often encoded in a relative manner, the configuration space has to be encoded absolute. E.g., a robot could have an action \textit{turn left} which would result in a different orientation depending on its previous orientation. For the configuration, an absolute encoding is necessary, e.g., the orientation as a cardinal point. This can result in an implicit integration of the state in the configuration, e.g., when the current orientation is part of the state. Furthermore, the local reward is a concept adopted from the reward in the RL domain, but the difference is the focus on the locality. E.g., in a case where the local reward of each agent is equal (as in a fully cooperative game), the influence detection can still be applied but might be of limited use since it is only possible to measure which agent has influence on the overall result. This result can still be interesting but, in this article, the focus is on studying systems where the notion of a local reward is matched.

\subsection{Example Applications}\label{sec:200-system-model_example-applications}
In the following, we introduce several example applications for illustration purposes. 
We start with two elementary use cases that allow for a simple description and introduction in the appearing challenges. Afterwards, smart camera networks are introduced, which define a real-world application often used to demonstrate SASO concepts.

%
%
\subsubsection{Collaborative Box Manipulation}\label{sec:200-system-model_example-applications_toy-problems}
%
The collaborative box manipulation serves as first, rather simple scenario where mutual influences occur. It includes two robots and a heavy box. The robots can be configured to push or pull the box, but none of them is able to move the box alone since it is too heavy. Therefore, the robots have to cooperate, i.e., both push or pull the box. This means that the configuration space is $\{PUSH,PULL\}$. The robots receive a local reward that is $1$ if the box moves forward or $0$ otherwise. Obviously, the configuration of one robot has an influence on the success of the other.

\subsubsection{Two-Man Saw}
The two-man saw scenario is a second example which is quite similar to the previous one. Again, it includes two robots, but, in comparison to the box example, a two-man saw is operated instead of pushing a box. The robots can be configured to push or pull the two-man saw (which results in the same configuration space as before), but they cannot handle the saw alone. Therefore, the robots have to cooperate, i.e., one has to push and the other one has to pull the saw (and vice versa). The robots receive a local reward that is $1$ if the saw moves or $0$ otherwise. Again, a human can easily recognize that the robots influence each other.

%
%
\subsubsection{Smart Camera Network}\label{sec:200-system-model_example-applications_smart-cameras}
\begin{figure}[t]
  \centering
    \includegraphics[width=3cm]{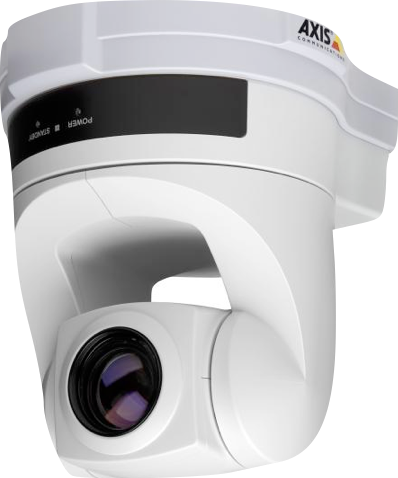}
  \caption{An example smart camera as used in our lab.}
  \label{fig:system-model:smart-cameras}
\end{figure}
Smart cameras (SCs) possess a build-in computation unit that can be utilized for several tasks such as image processing, object localization, and object tracking. Most SCs have pan, tilt, and zoom capabilities and the computation unit is used to determine beneficial alignments for the camera. Beyond, SCs are equipped with wired or wireless communication devices that allow for communication with neighboring cameras, and, therefore, form a SC network.
Such SCs can be utilized to achieve different goals, including the tracking of objects, the identification of new objects or the 3D reconstruction of objects. In order to apply the methodology of detection of mutual influences to SCs, a local reward is required. Obviously, this reward function heavily depends on the chosen goal. Regarding the configuration,, we consider the three adaptable parameters of the camera's alignment (pan, tilt, zoom).
In order to cover different classes of possible mutual influences in SASO systems several goals can be considered:
\begin{itemize}
	\item the maximization of the observed space, 
	\item the coverage of fast changing environments \cite{RudolphETH2014}, 
	\item the detection of previously unknown objects \cite{RudolphTSHWH2015} and 
	\item the 3D-reconstruction of objects \cite{RudolphTSH2015}.
\end{itemize}
%
%

Concluding the former discussion and mapping this application to the model of the target system, it is a realistic assumption that the cameras’ configuration space is composed of their pan, tilt and zoom configuration. Allowing a pan between 0 and 360 degree, a tilt between 0 and 90 degree and a zoom between 12 and 18, this would be $[0,360)\times[0,90]\times[12,18]$ for each camera. As stated before, the configuration spaces of the cameras are identical in structure, but each camera can assume an individual configuration. For the local reward we can chose one of the introduced functions.

%
%
\section{Related Work}\label{sec:300-detection-of-influences_related-work}
In the following, the related work in the area of influence formalization and quantification is presented. Further related work can be found in Section~\ref{sec:300-detection-of-influences_general-methodology_dependency-measures} where several dependency measures are introduced. 

In literature, several approaches can be found that try to formalize the mutual influences. Most of these approaches focus on the influence through direct or indirect interactions. For instance, a model for interactions is proposed by Keil et al.~\cite{KeilG2003}, but a method to detect the implicit interactions is not provided. Another common approach is to use \textit{stit} logic for modeling the interactions in multi-agent systems~\mbox{\cite{LogieHW2008,LogieHW2010,Broersen2010}}. The focus of these works is on the system specification and verification and therefore differs fundamentally from the focus of our work where the goal is an adaption at runtime.\\

Multi-agent reinforcement learning (MARL) is an active field of research \cite{StoneV2000,BusoniuBD2008}. 
An overview and a useful taxonomy has been introduced by Busoni et al.~\cite{BusoniuBD2008}, for instance.
Following their taxonomy, the approach presented here can be useful in fully competitive and mixed games (each static or dynamic). It can be used in fully cooperative games as well, especially if the global payoff is a function of the local payoffs of the agents (cf. the smart camera application). As stated by the authors, the complexity resulting from coordination is a major issue in MARL systems. The influence detection mechanism presented here can keep this complexity to a minimum since it allows to only coordinate with relevant partners.

In the following, we outline the related work from the MARL domain. Kok et al.~\cite{KokSV2003} presented work that is based on so-called coordination graphs and an approach to solve the global coordination problem on a local basis if it is possible to decompose the global payoff function into a sum of local payoff functions. A restriction of this method is that it relies on inference rules that are hand-crafted. Furthermore, they focus on discrete state variables which is not the case in the approach presented in this article. But, since they assume a given graph, the influence detection could be used to infer such a graph based on the roles of the agents.

An extension to this approach has been presented in~\cite{KokHBV2005}. There, similar to the influence-based approach, the coordination graph is inferred at runtime creating a transition from independent learners to coordinated action selection. It is based on a t-test between the maximally possible expected reward when the agents act in common and the expected reward from independent decisions. The approach is for general sum games and is limited to discrete state and action spaces, which means that it cannot be simply adopted to a continuous case. Furthermore, the work only shows the applicability for a Q-learning-based approach and there is no trivial way to make it applicable to a wide range of algorithms. Methodologically, a major difference here is that the method is focused on finding the states in which agents should collaborate. This requires that all agents are always willing and able to cooperate.

DeHauwere et al.~\cite{DeHauwereVN2009} demonstrated a specialized solution for mazes with two robots. They used a generalized learning automaton that uses the distance to another robot to learn how to avoid a collision by identifying states in which they have to coordinate. While this approach might be adapted for other tasks it would be necessary to hand-craft the states of the learning algorithm each time.

Later, the authors presented an article focusing on a method to generalize the learned behavior for a single state over several states. However, the presented approach needs to identify the states where coordination is necessary~\cite{DeHauwereVN2010}. Their method is similar to Kok's. However, they assume that the agents have already learned an optimal policy if acting alone since this is necessary to find states that need coordination which renders it unusable for learning at runtime. Furthermore, the work is focused on sparse interaction, i.e., it identifies the states for cooperation and not the systems. This means the approach is not appropriate to find useful collaborators from a set of agents. Furthermore, the approach is limited to discrete state and action sets. Another difference to the approach presented here is that it relies on the Kolmogorov–Smirnov test used as a goodness-of-fit test, i.e., it tests if the distribution of points fits a model. However, this approach depends on a model, which is not always present, e.g., if not Q-learning is used but a policy-based RL approach, which is contrary to the requirement of independence from the control mechanism of the system.

The above approach has been extended to solve delayed coordination problems~\cite{DeHauwereVN2011}. However, the limitation regarding the discrete state and action sets persists. Furthermore, the approach is still not independent from the control algorithm.

Lanctot et al.~\cite{Lanctot2017} \textcolor{black}{proposed a measure called joint policy correlation. This has been applied prototypically to a two agent laser tag scenario. It is based on the repetition of the same scenario with several seeds which leads to different strategies of the agents. Afterwards, a matrix is formed that compares the average rewards of the agents against the agents from the other repetitions. The values against the initial opponent and the other opponents are then aggregated to create a measure that allows to see how much an agent has overfitted to its initial opponent. The goal of the work is very different from the one presented here since it wants to create a measure on how much an agent overfits to the behavior of other agents. A measure of influence can not be directly derived from this approach. Furthermore, in this article, we focus on a runtime learning approach which is not possible if the experiment has to be rerun several times.}\\

The approach presented in this article has similarities with feature selection methods. These are categorized as \textit{filter}, \textit{wrapper} or \textit{embedded} methods. Wrapper methods are not applicable here since they require multiple repetitions of the learning tasks which is contrary to the goal of learning at runtime. In the field of filter methods, the minimum-redundancy-maximum-relevance~\cite{PengLD2005} (MRMR) approach is frequently used. However, it relies on the elimination of redundancies between the features which is not desired here since it reduces the information one gets from the influence detection. Furthermore, it comes at an extensive computational cost compared to the methods proposed in this article. 

Regarding the embedded methods, the most prominent instance is the regularized least-squares policy~\cite{FarahmandGSM2008,KolterN2009,LiuLW2015}. This and similar methods are tied to a specific learning algorithm and therefore breach the restriction of an applicability independent from the control algorithm.\\
\textcolor{black}{Furthermore, there are attempts to combine self-organizing systems with learning algorithms, e.g.,} Boes et al.~\cite{Boes2017} \textcolor{black}{introduced a MAS for the control of technical systems that has SASO characteristics. The method is called \textit{Escher} and aims to deconstruct complex control problems by using an acyclic network of interacting agents that are able to learn from experiences. The algorithm has been applied to toy problems and an engine calibration. The approach differs from the one presented here since they created a MAS to control a single technical system while here we focus on several systems that are autonomous and have to adapt to the behaviour of the other systems. But, we believe that an application of the influence detection to the individual controller agents of Escher can be helpful if the different inputs of the controlled system have to be coordinated to reach optimal results.}\\

Concluding the related work, there is no algorithm that allows to identify subsystems with continuous and discrete states and actions which influence other subsystems to find useful collaborators and is independent from the control algorithm.

%
%
\section{Mutual Influences}\label{sec:300-detection-of-influences}
In this section, we present our basic method for the detection of influences among autonomous SASO systems. 
{\color{black}This method is a refined version of preliminary work that can be found in \cite{RudolphHTH2016,RudolphTH2016,RudolphTSH2015} and necessary to understand the subsequent variants presented in the following sections. Especially, the refinement brings the method in a wider scope by defining a general workflow and applies it to new scenarios. Besides extending the basic method in the following sections, we simultaneously incorporate more complex tasks, i.e. the detection at runtime and the possibilities of an intelligent SASO system to adapt its own behaviour to detected influences.}

%
%
\subsection{Methodology for Detection}\label{sec:300-detection-of-influences_general-methodology}
Based on the described target systems, we define the methodology for mutual influence measurement. The goal is to identify those components of the configuration of the other systems (i.e., value range and considered variables) that have influence on the system itself. After the identification of influencing configuration components, they can be addressed by a designer, e.g.\ by considering them in control algorithms, or by a self-adapting system itself, e.g.\ using a learning algorithm.

In general, we are interested in the question whether a system as a whole is influencing other systems. However, to be more precise in the description of the influence, we want to detect those parameters where the optimal configuration values are somehow influenced by the current settings of the other systems. The basic idea of the following approach is to make use of stochastic dependency measures that estimate associations and relations between the configuration components of a system and the reward of a second system. The basic method assumes that the mutual influence between the systems is instantaneous, meaning that the reward of the system reacts to the configurations of the other systems in the same time step. However, the approach can be extended in order to detect delayed influences as well by measuring the dependency between the configuration components and a later reward~\cite{RudolphHTH2017}. 

In general, dependency measures are designed to find correlations between two random variables. In the following, we model the configurations of distributed systems as such random variables, which reflects the system model where we face autonomous systems whose actions can be uncertain due to non-deterministic behavior or incomplete information, for instance.

\begin{figure}[]
    \centering
    \resizebox{4cm}{!}{%
    \begin{tikzpicture}[node distance=1.5cm]

        \node (observation) [process] {Observation};
        \node (distribution) [process, below of=observation] {Distribution};
        \node (estimation) [process, below of=distribution] {Estimation};
        \node (evaluation) [process, below of=estimation] {Evaluation};
        \node (adaptation) [process, below of=evaluation] {Adaptation};

        \draw [arrow] (observation) -- (distribution);
        \draw [arrow] (distribution) -- (estimation);
        \draw [arrow] (estimation) -- (evaluation);
        \draw [arrow] (evaluation) -- (adaptation);
        \draw [arrow,dashed] (adaptation) --++ (+3,0) |- (observation);
    \end{tikzpicture}
    }
    \caption{The general workflow of influence detection.}
    \label{fig:influence-detection:workflow}
\end{figure}
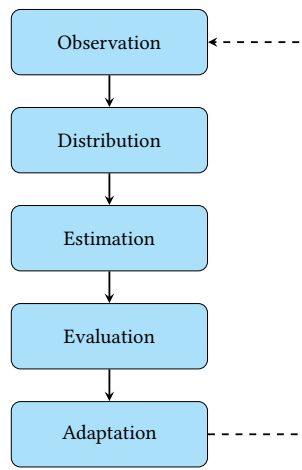

Before introducing relevant candidates for the dependency measures, we outline the general workflow (cf. Figure~\ref{fig:influence-detection:workflow}) for an influence detection for each subsystem in the overall system.

\begin{enumerate}
    \item \emph{Observation:} Continuously observe the configuration and estimate the goal achievement of a productive system, e.g. the pan, tilt, and zoom of a camera and the corresponding reward in terms of the given goals. These observations will typically be done by the system itself but they could as well come from an external entity.
    
    \item \emph{Distribution:} Gather configurations from other systems and provide your own one to them. 
    
    \item \emph{Estimation:} Estimate the dependency value by relating the own reward to the configurations of the other systems. The basic idea of the algorithm is to make use of stochastic dependency measures that estimate associations between the reward of a system $A$ and the configuration components of a second system $B$. These dependency measures are designed to identify correlations between two random variables $X$ and $Y$. The reward of $A$ is identified with a random variable $X$ and the configuration of entity $B$ with a random variable $Y$. This mapping implies that if the association between $X$ and $Y$ is high, we also have a high influence of $B$ on $A$, since it reflects that the configurations of $B$ \textit{matter} for the reward of $A$. Vice versa, if the association is low, we do not see an influence. There are several dependency measures available which might be suitable depending on the application. An overview for the most interesting measures is given in the following. Furthermore, it might be necessary to take other configuration parts into account. These are most likely the configuration parts of the influenced agent, i.e., the configuration of system $A$ is relevant for the measurement of the correlation between the reward of system $A$ the configuration of system $B$. In this case, we condition the calculation with the configuration of $B$. This means that we calculate the dependency of the reward and the configuration for each configuration of $B$ separately in the discrete case. In the continuous case, we split the configuration in two (or possibly more) parts and calculate the dependency separately. For a more detailed explanation of the estimation please see \cite{RudolphHTH2016,RudolphTSH2015}.
    
    \item \emph{Evaluation:} This step compares the values calculated for the different (other) systems and their configuration components. There are several possibilities to do so: The first approach would be to compare the influence values to a fixed threshold, which is a valid option but can be difficult since an appropriate threshold might not be available for each application. Second, the influence values can be compared on a relative basis between the systems. This leads to a ranking of systems according to their influence. A third possibility is to simulate an independent system with the same configuration space and compare the \enquote{artificial} value to the real values of the other systems. This allows to decide on a basis that only includes one other system. Additionally, it can be useful to calculate the p-value, i.e. a value that allows to estimate how likely the given outcome under the assumption of independence is.
    
    \item \emph{Adaptation:} This step addresses the influences in the control strategy, which can take various forms: Especially when applied at design-time, the designer can decide on a case-by-case basis. For a self-adaptive solution, we propose to use a learning algorithm that includes the configuration of the influencing systems in the situation description during runtime.
\end{enumerate}

This process can be used either at design-time with a prototype or continuously at runtime. For the design-time variant, the system runs for a certain time while the configurations and rewards are logged. Afterwards, the calculation and decision process starts. During runtime, the steps have to be considered as a loop that runs continuously but one can still decide on how long samples are gathered since a distribution and recalculation may not be justified for each newly gathered sample.

%
%
\paragraph{\textbf{Discussion of Dependency Measures}}\label{sec:300-detection-of-influences_general-methodology_dependency-measures}

{\color{black}As mentioned before, our method relies on the utilization of basic dependency measures for the influence detection. As in \cite{RudolphHTH2016}, we briefly introduce the most prominent measures and consider their advantages and drawbacks in the following list. The contained measures are then analyzed in the following evaluation to derive statements about their possible applicability for the ranking task in our method.}

\begin{itemize}
    \item \emph{Pearson correlation:} The probably most prominent instance is the Pearson correlation coefficient -- sometimes just called \enquote{correlation coefficient} \cite{Pearson1895}. The main advantages are its simple implementation and its fast calculation. In the context, the major drawback is that only linear correlations can be detected, i.e., it can fail in case of more complex dependencies. Moreover, it is necessary to calculate the distance between realisations of the random variable which might make it not well suited for some problems. It assumes values between $-1$ and $1$, where $-1$ indicates a perfect negative linear correlation and $1$ a perfect positive correlation. $0$ means that there is no linear correlation. When comparing the influence of different systems, it should be considered to use the absolute values. The Pearson correlation coefficient $r$ is defined as:
    	
    \begin{equation}
        r = \frac{\sum_{i=1}^{n}(x_{i}-\overline{x})(y_{i}-\overline{y})}{\sqrt{\sum_{i=1}^{n}(x_{i}-\overline{x})^{2}}\sqrt{\sum_{i=1}^{n}(y_{i}-\overline{y})^{2}}}~,
       \label{equ:basic_methods:influence_detection:7_4_1}
    \end{equation}
    
    where the $x_{i}$ and $y_{i}$ are the gathered samples, i.e. the configuration components and the reward. $n$ is the number of samples and $\overline{x},\overline{y}$ denote the mean values of the random variables.
    \item \emph{Kendall rank correlation:} Another measure that is based on calculating the ranks of the gathered samples has been introduced by Kendall~\cite{Kendall1938}. The measure can be computed rather fast. It can detect monotone dependencies, which is better than just linear dependencies but still can be not sufficient for many applications. It assumes values between $-1$ and $1$; where $-1$ indicates a perfect monotone declining relationship and $1$ a perfect monotone increasing relationship. For independent variables, a value around $0$ should be considered. The Kendall rank correlation $\tau$ is calculated using: 
    
    \begin{equation}
        \tau = \frac{(\#~concordantPairs)-(\#~discordantPairs)}{n(n-1)/2}~.
        \label{equ:basic_methods:influence_detection:7_4_2}
    \end{equation}
    
    A concordant pair are two samples ($x_{i}, y_{i}$), ($x_{j}, y_{j}$) where the ranks of the elements agree, i.e. if $x_{i}>x_{j}$ then $y_{i}>y_{j}$ or if $x_{i}<x_{j}$ then $y_{i}<y_{j}$. The opposite of a concordant pair is called discordant. Cases with $x_{i}=x_{j}$ or $y_{i}=y_{j}$ are neither discordant nor concordant and are not handled in the basic variant.
    
    \item \emph{Spearman rank correlation:} The rank correlation after Spearman is similar to Kendall's since it is based on ranks, too. Here, the ranks of the samples are calculated and instead of a comparison between the samples the Pearsons correlation coefficient is calculated on the ranks. Therefore, the values range from $-1$ to $1$ with $0$ meaning that no dependency has been detected. The Spearman correlation is calculated using:
    
    \begin{equation}
        \rho = \frac{\sum_{i=1}^{n}(rg(x_{i})-\overline{rg_{x}})(rg(y_{i})-\overline{rg_{y}})}{\sqrt{\sum_{i=1}^{n}(rg(x_{i})-\overline{rg_{x}})^{2}}\sqrt{\sum_{i=1}^{n}(rg(y_{i})-\overline{rg_{y}})^{2}}}~,
        \label{equ:basic_methods:influence_detection:7_4_3}
    \end{equation}
    
    where $x_i$ and $y_i$ denote the samples and $n$ the number of samples. $rg(x)$ is short for the rank of $x$, i.e. the position of $x$ if all samples are ordered by their value. $\overline{rg_x}$ is the average rank of the samples $x_i$ and $\overline{rg_x}$ for $y_i$, respectively. 
    
    \item \emph{Distance covariance:} This measure is an extension to the Pearson correlation that takes the distance between the samples into account. Since these distances have to be calculated for the entire sample set, it is not suitable for an online calculation. The advantage of this method is that it is not limited to linear dependencies but can find all types of dependencies \cite{SzekelyRB2007}. The distance covariance is calculated as follows: We first derive the Euclidean distances $a_{j,k} = \Vert x_{j} - x_{k} \Vert$ and $b_{j,k} = \Vert y_{j} - y_{k} \Vert$. Afterwards, we calculate $A_{j,k}:=a_{j,k}-\overline{a_{j.}}-\overline{a_{.k}}+\overline{a_{..}}$ and $B_{j,k}:=b_{j,k}-\overline{b_{j.}}-\overline{b_{.k}}+\overline{b_{..}}$ where $\overline{a_{j.}}$ is the mean of the $j$-th row, $\overline{a_{.k}}$ is the mean of the $k$-th column and $\overline{a_{..}}$ is the mean of the whole matrix:
    
    \begin{equation}
        \frac{1}{n^{2}}\sum_{j=1}^{n}\sum_{k=1}^{n}A_{j,k}B_{j,k}
        \label{equ:basic_methods:influence_detection:7_4_4}
    \end{equation}
    
    \item \emph{Mutual information:} A quite different approach has been taken by Shannon in the context of information theory \cite{ShannonW1949}. The basic variant can be used for discrete random variables and can be calculated online. Furthermore, all types of dependencies can be found. A disadvantage is that the maximal possible value depends on the structure of the random variables, i.e. the values can be normalized between $0$ and $1$, but the comparability to other variables with different structures might be limited. The mutual information is defined as:
    
    \begin{equation}
        I(X;Y): = \sum_{x \in X}\sum_{y \in Y}p(x,y)\ld \left(\frac{p(x,y)}{p(x)p(y)} \right)
        \label{equ:basic_methods:influence_detection:7_4_5}
    \end{equation}
    
    where $p(x,y)$ is the joint probability of the events $x$ and $y$ and $p(x)$ and $p(y)$ are the marginal probabilities of $x$ and $y$. The probabilities can be easily approximated using a frequency counting (i.e., a maximum likelihood approach based on discrete finite random variables). There exists also a continuous variant of the mutual information measure. Here, the calculation is more complex since the estimation from samples needs more advanced techniques. The most common approach is based on a $k$-nearest neighbor method \cite{KraskovSG2004}. Another approach is the manual binning of the values in order to use the discrete version, which can bear problems because the results might depend highly on the chosen binning parameters. An automatic binning can be done with the maximal information coefficient (see below).
    \item \emph{Maximal information coefficient:} This measure is an extension to the mutual information for real-valued data. It automatically calculates the binning that results in the maximal mutual information for a given data set. A drawback is that there is no method for the online calculation. The basic formula is:
    
    \begin{equation}
        MIC(X,Y): = \max_{n_{x}n_{y}<B} \frac{I(X;Y)}{\log(\min(n_{x},n_{y}))}
        \label{equ:basic_methods:influence_detection:7_4_6}
    \end{equation}
    
    where $n_{x}$ and $n_{y}$ denote the number of bins for $X$ and $Y$. This means that the number of bins is limited by a threshold $B$ that is by default determined depending on the sample size. The denominator serves for the normalization of the value. However, it is computationally expensive to determine the values for all possible binnings. Therefore, a heuristic is introduced. For details please see \cite{ReshefRFGMTLMS2011}.
\end{itemize}

It should be mentioned that the choice of a dependency measure does not have to be exclusive. It can be useful to calculate several measures in parallel or to perform an iterative process, which initially uses computationally light-weighted measures and adds more powerful and computationally expensive methods only if necessary.

\subsection{Evaluation}\label{sec:300-detection-of-influences_general-methodology_evaluation}
{\color{black}Initially, we aim at investigating which measurement from the previous subsection is most suitable for the detection of mutual influences. Therefore, we consider the three initially introduced application scenarios that are characterised by increasing complexity. }

%
\subsubsection{Collaborative Box Manipulation}
In this scenario, two robots have the task of pushing a box. Each of them can either $PUSH$ or $PULL$ the box, but the box only moves if both of them $PUSH$. This leads to a reward of 1 while all other combinations lead to a reward of 0. Since some of the dependency measures do not allow for categorical random variables, such as the here used $PUSH$ and $PULL$, the configuration has been mapped to the numbers 0 and 1.

\begin{figure}[]
	\centering
	\begin{subfigure}{0.7\textwidth}
		\centering
		\includegraphics[width=\textwidth]{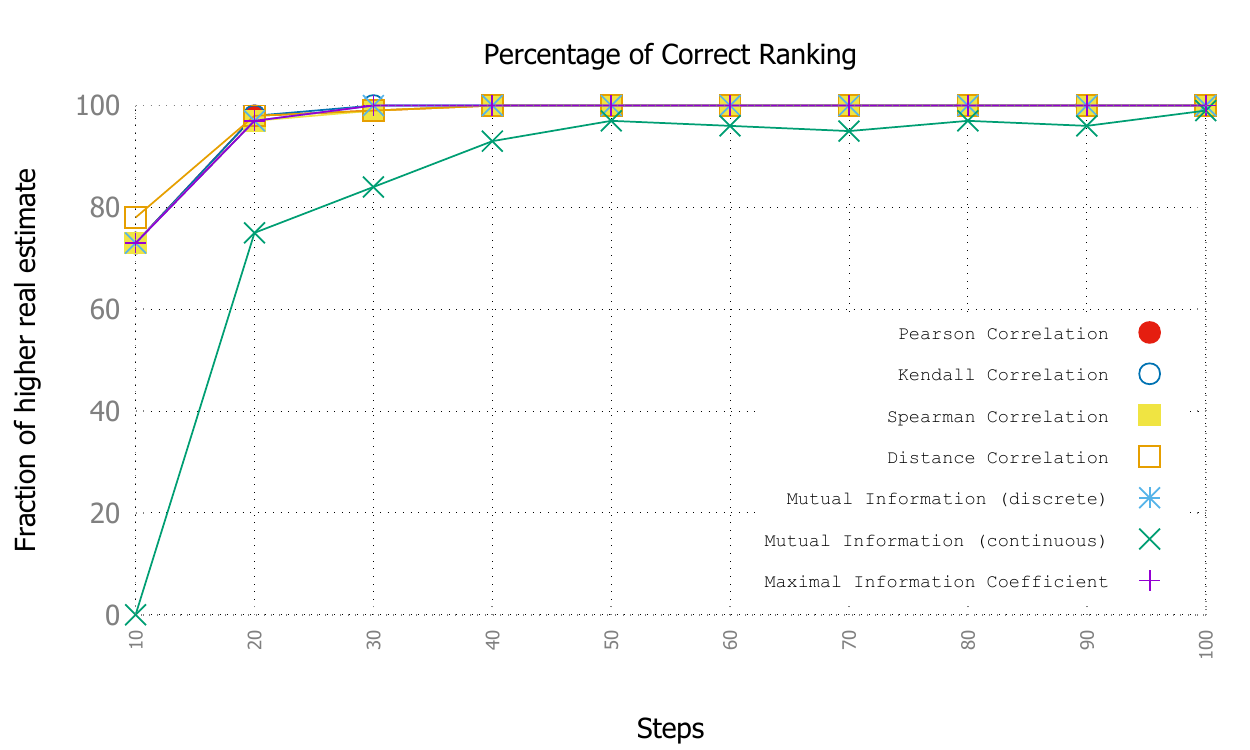}
		\caption{An overview of the full 100 steps evaluated.}		
	\end{subfigure}
	
	\begin{subfigure}{0.7\textwidth}
		\centering
		\includegraphics[width=\textwidth]{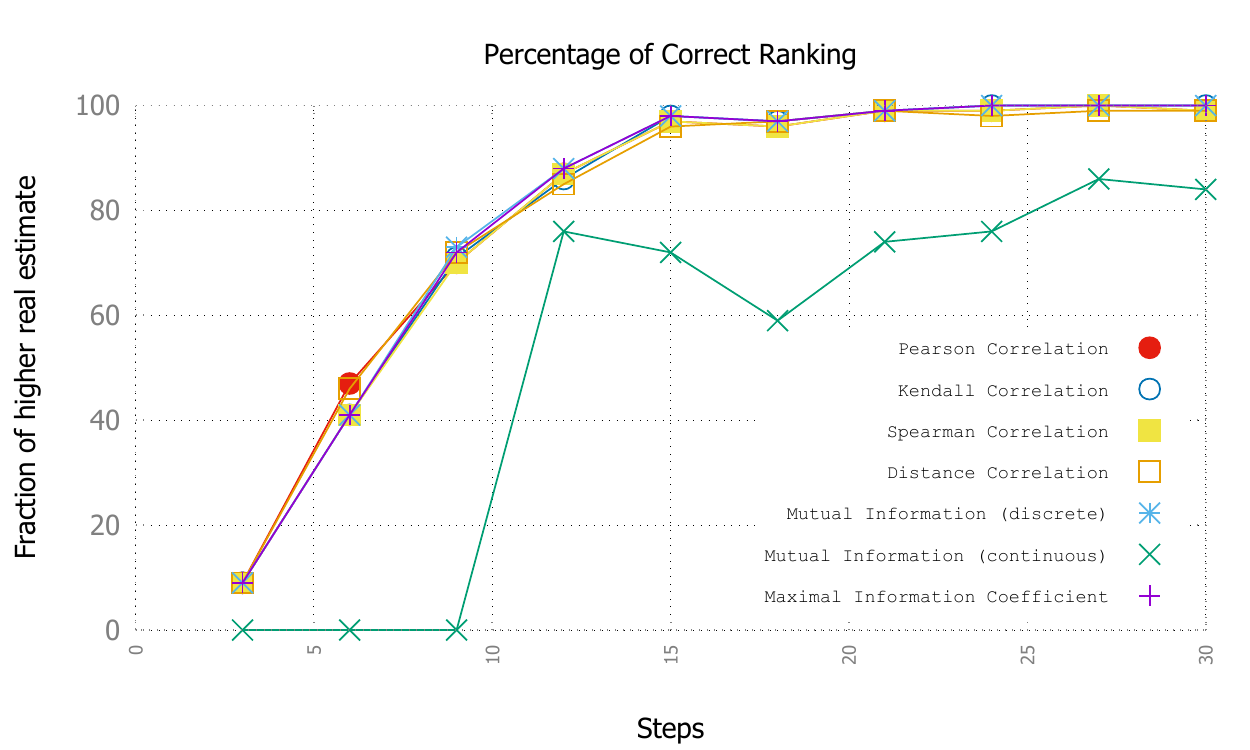}
		\caption{A more detailed view on the first 30 steps.}
	\end{subfigure}
	\caption[The results for the collaborative box manipulation.]{The results for the collaborative box manipulation. The graphs show the fraction of 100 runs in which the influence of robot B is detected higher as the influence of the notional robot.}\label{fig:300-detection-of-influences_general-method_eval_box-oneEstimator}
\end{figure}
\afterpage{\clearpage}

For the evaluation of this application, we conducted 100 experiments and measured the influence using each of the previously introduced dependency measures. For each run and each measure, two values have been calculated. The first one is the influence of robot $B$ on robot $A$. The second one is the influence of a notional robot. This notional robot has the same capabilities as the real robot $B$, but his actions do not influence robot $A$. The value is calculated to assess how reliable the detection of the influencing robots is. This can be done by comparing in which fraction of the runs robot $B$ has been found more influential than the notional robot.

The results are depicted in Figure \ref{fig:300-detection-of-influences_general-method_eval_box-oneEstimator}. There, we see that most of the measures perform similar in this scenario. Except for the continuous MI approximated with the Kraskov method 40 steps are sufficient to distinguish the influencing robot from a non-influencing in each of the 100 runs. The vast majority is already correctly detected after 15 steps. The continuous MI eventually finds the influence in each run. However, the detection speed is rather slow. This is due to the fact that the continuous MI has problems with the discrete values that are assumed in the problem. Furthermore, the continuous MI has not been calculated for less than ten steps since a minimum number of samples is required. Therefore, the first three data points should be disregarded.

Concluding the results, the influence in this elementary use case can be detected quite easily with the previously introduced method. Even though the continuous mutual information shows a little slower detection, the selected measure for the task is not too important since the detection is over all quite fast.

%
%
\subsubsection{Two-man Saw}
As a second example, we evaluate the conditioned measurement in an elementary use case, the two-man saw. 
The example is inspired by two robots that operate a saw that only moves if both of them move it in the right direction, i.e., one $PUSH$es and the other $PULL$s or vice versa depending on the current position of the saw. If the saw moves each robot gets a reward of 1 otherwise 0. Again, the categorical configurations $PUSH$ and $PULL$ have been mapped on numerical values to make all dependency measures applicable.

\begin{figure}	
	\centering
	\begin{subfigure}[t]{0.7\textwidth}
		\centering
		\includegraphics[width=\textwidth]{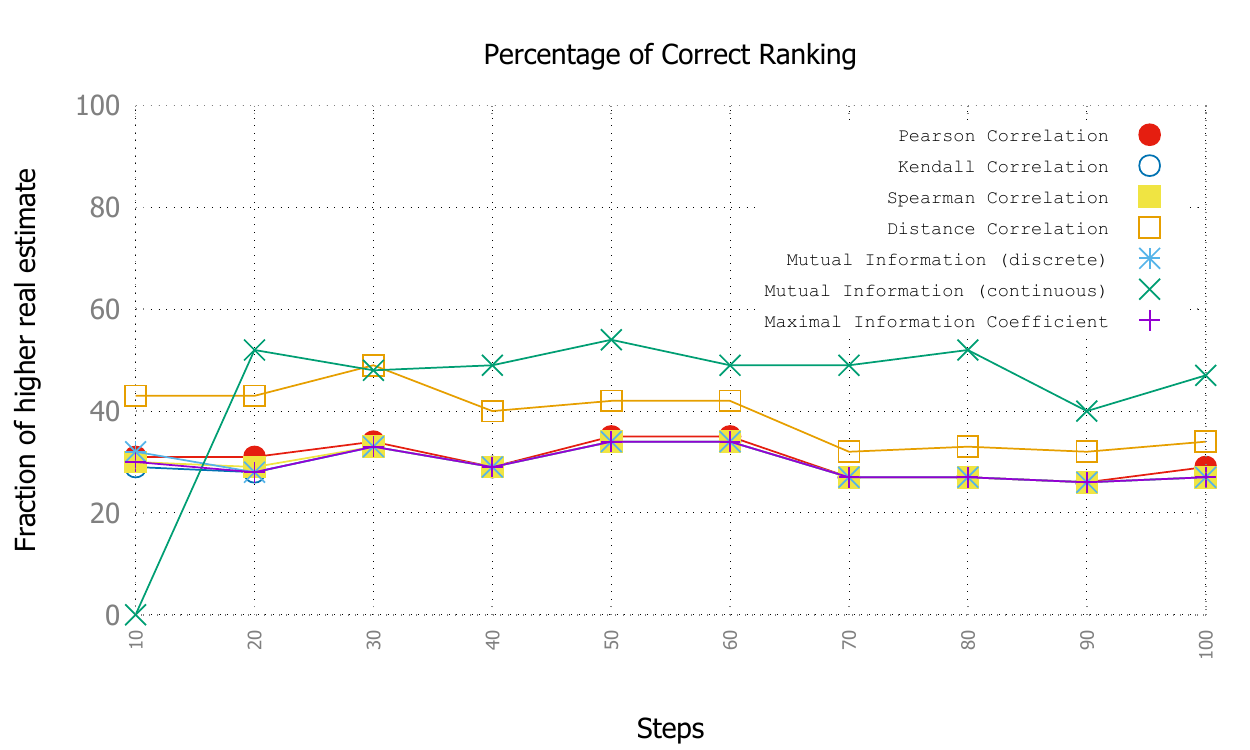}
		\caption{The detection if a single estimator is used and the own configuration is not considered.}		
	\end{subfigure}
	\\
	\begin{subfigure}[t]{0.7\textwidth}
		\centering
		\includegraphics[width=\textwidth]{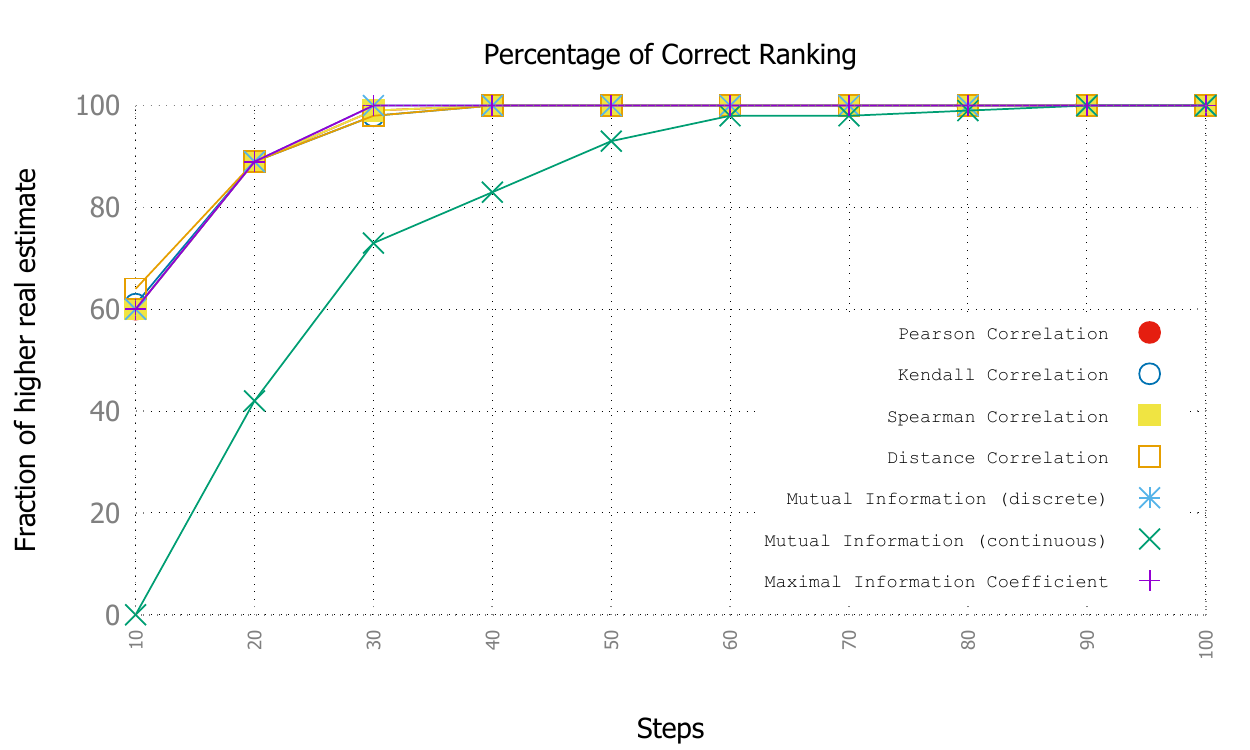}
		\caption{The detection if the own configuration is considered, i.e., there is one estimator for each of the configurations (Push and Pull).}
	\end{subfigure}
	\caption{The results for the two-man saw use case.}\label{fig:300-detection-of-influences_other-config_eval_saw}
\end{figure}

The results are depicted in Figure \ref{fig:300-detection-of-influences_other-config_eval_saw}. As for the collaborative box manipulation, the actual influence has been calculated using the seven dependency measures and compared with a measurement of a notional robot that assumes uniformly distributed random configurations but has no influence on the actual outcome of the experiment. The figure shows in how many of the 100 independent runs the influencing robot has been found to be more influential than the notional (not-influencing) robot. The first graph shows the result for the method used in the previous evaluations, i.e., without a consideration of other configurations. The detection is between $30\%$ and $50\%$ which is below the expected value of $50\%$. This due to the fact that the run will only be counted as correctly detected if the value is higher but not if both values are equal which is quite often the case in this scenario. The second graph shows the results with the consideration of the other configuration, i.e., the calculation of the dependencies is conditioned under the configuration of the first robot. We see that influences are almost perfectly detected after 30 steps which is close to the result in the collaborative box manipulation. Also similar is the small weakness of the continuous MI measure which needs about 90 steps to catch up.

Concluding the evaluation of the two-man saw application, we have seen that even in rather simple applications the problem with other configurations can lead to influences that can not be detected with the basic variant of the influence detection. However, the improvement that includes the other configurations by conditioning the measurement shows very good results for such cases as well.

\afterpage{\clearpage}

%
%
\subsubsection{Smart Camera Network}
As a third application for the evaluation, we consider an example for the SC network domain. To recap briefly, SCs are surveillance cameras that are equipped with computational capabilities that can be used for several tasks including image processing. Furthermore, these cameras are interconnected via a network that allows them to exchange data and to coordinate. For this article, we stick to so-called PTZ cameras that allow for an automatic adjustment of the pan angle, tilt angle, and zoom of the camera. As previously described, there are several reasonable goals for a SC network. For this evaluation, we stick to the goal of a 3D-reconstruction of the observed targets. In particular, this means to observe the targets from different perspectives and cameras at the same time. To achieve this, the cameras get a reward of 1 for each object that is observed by at least two cameras in a time step.

\begin{figure}[t]
	\centering
	\includegraphics[width=8cm]{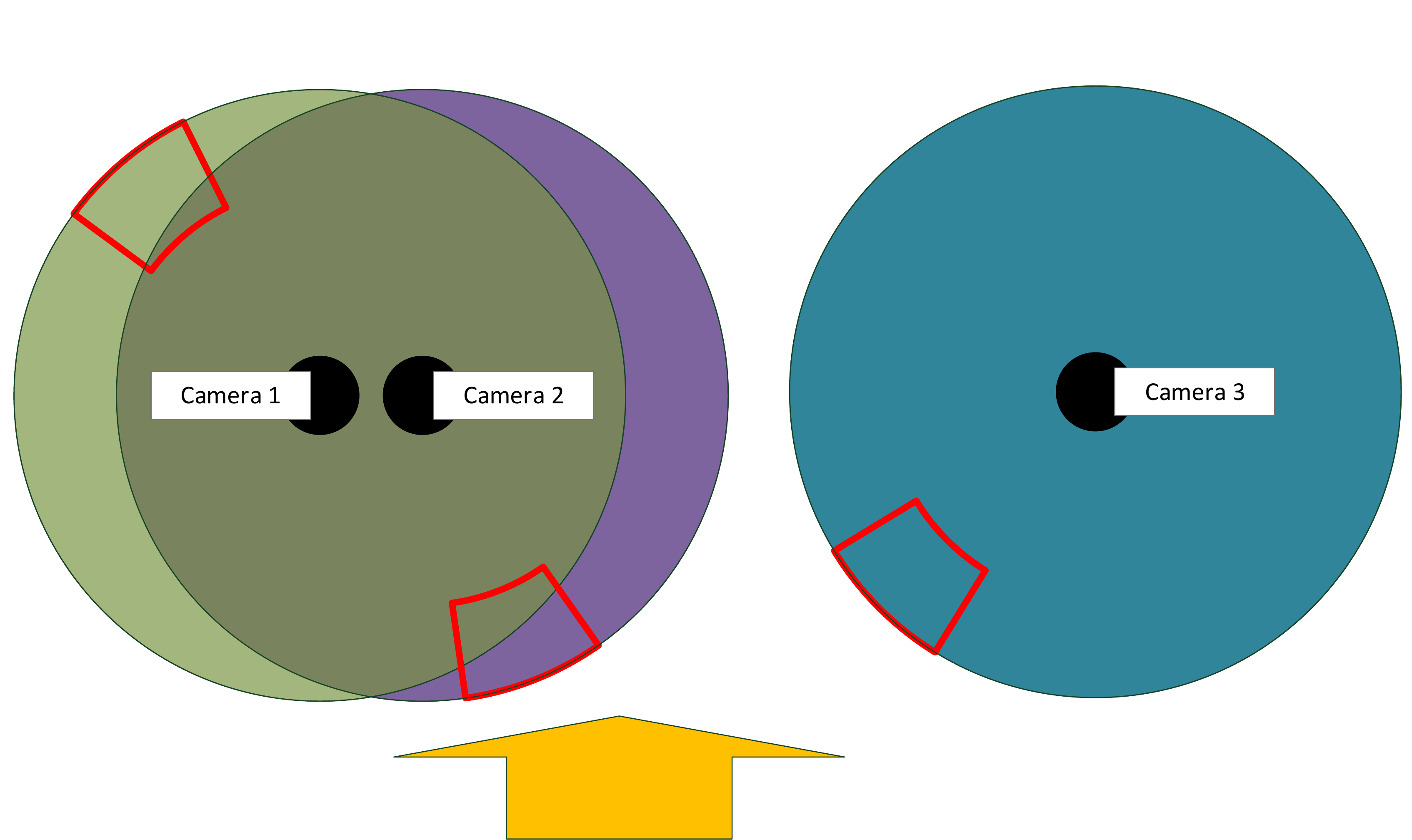}
	\caption[The smart camera scenario \textit{SCN 1}]{\textit{SCN 1}. A top-down view on a smart camera network. The black dots depict cameras surrounded by a circle that marks their potential observable area. The red shapes show the field of view for an exemplary PTZ configuration. The yellow arrow indicates from where and in which direction the objects of interest move.}
	\label{fig:eval:camera:testDetection4}
\end{figure}

A top-down view on scenario SCN 1 is depicted in Figure \ref{fig:eval:camera:testDetection4}. There, we see three black dots which represent one camera each. Around the dots, there are colored circles. Each of them represents the area that is potentially observable by one of the cameras if it chooses an according PTZ configuration. Furthermore, we see example areas that are currently observed marked my red lines. Camera 1 and Camera 2 share a common area which can be observed by both at the same time. In contrast, Camera 3 is isolated and does not share a common area with one of the others. The yellow arrow marks the entry point and direction for objects that move through the scene in the area between Camera 1 and Camera 2.

Ten independent runs of this scenario have been conducted in a Mason\footnote{Mason is a multi-agent simulation framework for Java~\cite{LukeCPSB2005}.} simulation, where each of the cameras assumes a uniformly sampled PTZ configuration in each time step. The pan angels are between $0$ and $360$ degree, the tilt angle between $120$ and $180$ degree, and the zoom between $12$ and $18$. In each of the runs, we compared the influence of Camera 2 on Camera 1 and the influence of Camera 3 on Camera 1. For each of the measures, there are three comparisons: one for the pan, one for the tilt, and one for the zoom of the cameras. It is expected that there is a clear trend towards a higher influence of Camera 2 over Camera 3 in general and especially for the pan and tilt since these configuration determine if it is possible to gather a reward for Camera 1 or not.

\begin{figure}[t]
	\centering
	\includegraphics[width=8cm]{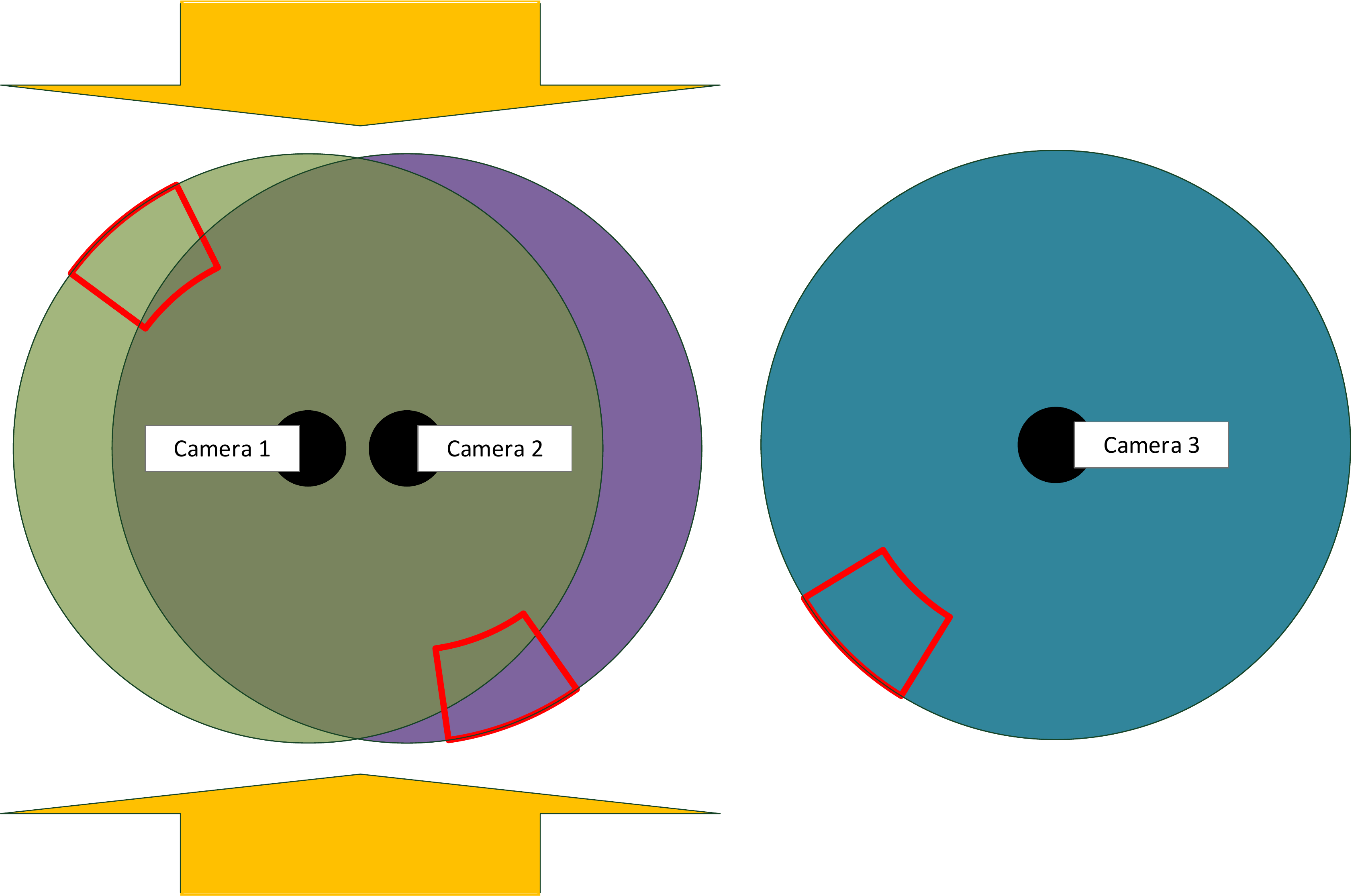}
	\caption[The smart camera scenario \textit{SCN 2}]{\textit{SCN 2}. A top-down view on a smart camera network. This scenario is only slightly changed from SCN 1 by adjusting the flow of targets represented by the yellow arrows. The black dots depict cameras surrounded by a circle that marks their potential observable area. The red shapes show the field of view for an exemplary PTZ configuration.}
	\label{fig:eval:camera:detectionStrong}
\end{figure}

\begin{figure}[]
	\centering
	\begin{subfigure}{\textwidth}
		\centering
		\includegraphics[height=0.25\textheight]{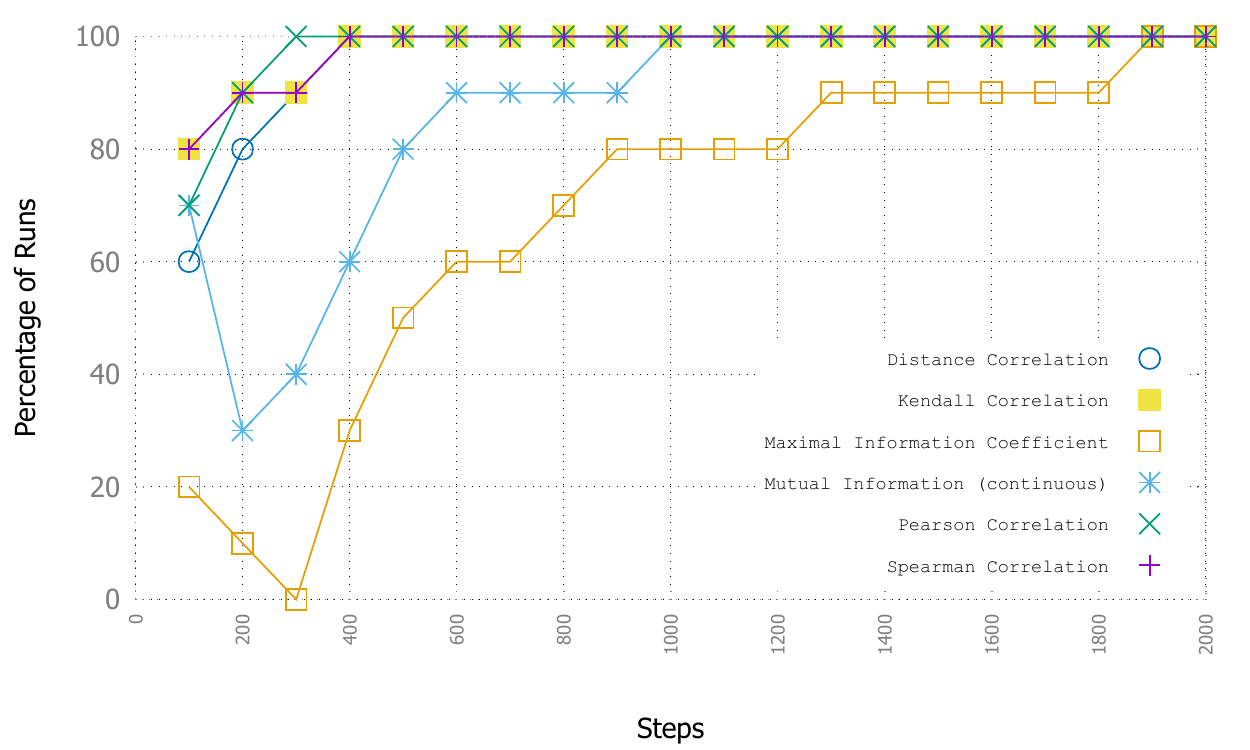}
		\caption{The results for the pan.}		
	\end{subfigure}
	\begin{subfigure}{\textwidth}
		\centering
		\includegraphics[height=0.25\textheight]{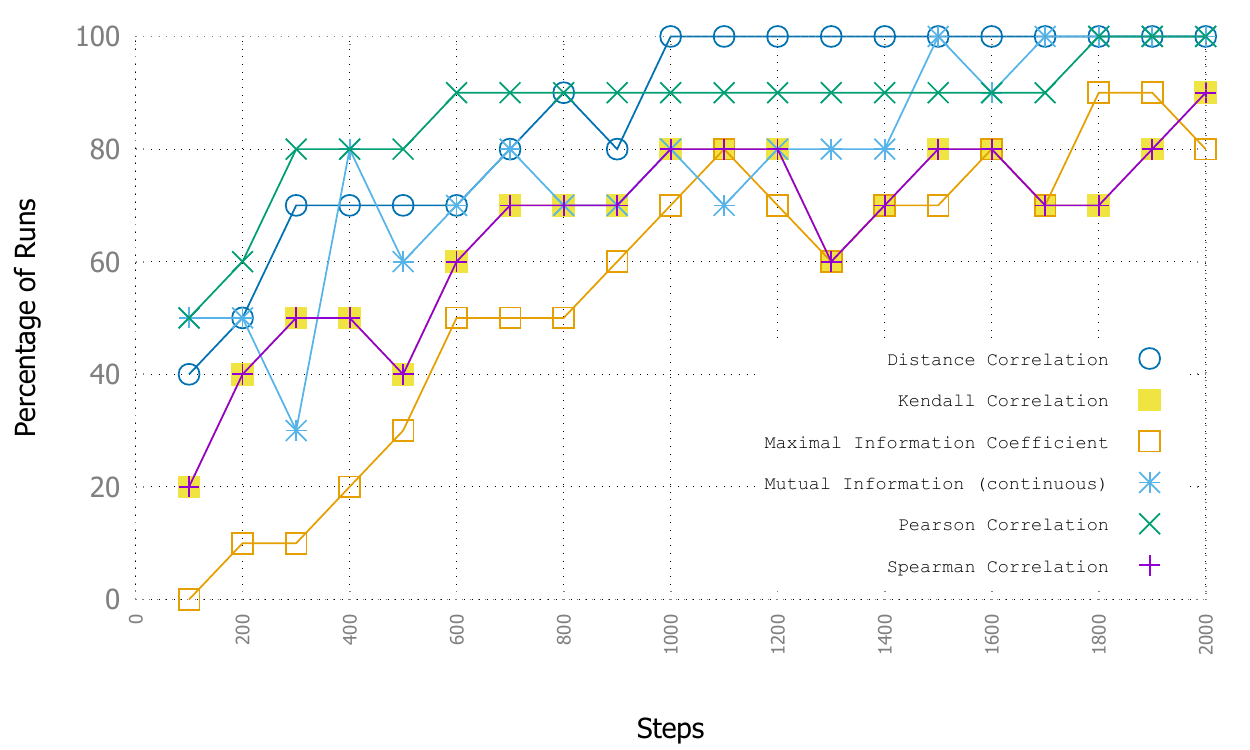}
		\caption{The results for the tilt.}
	\end{subfigure}
	\\
	\begin{subfigure}{\textwidth}
		\centering
		\includegraphics[height=0.25\textheight]{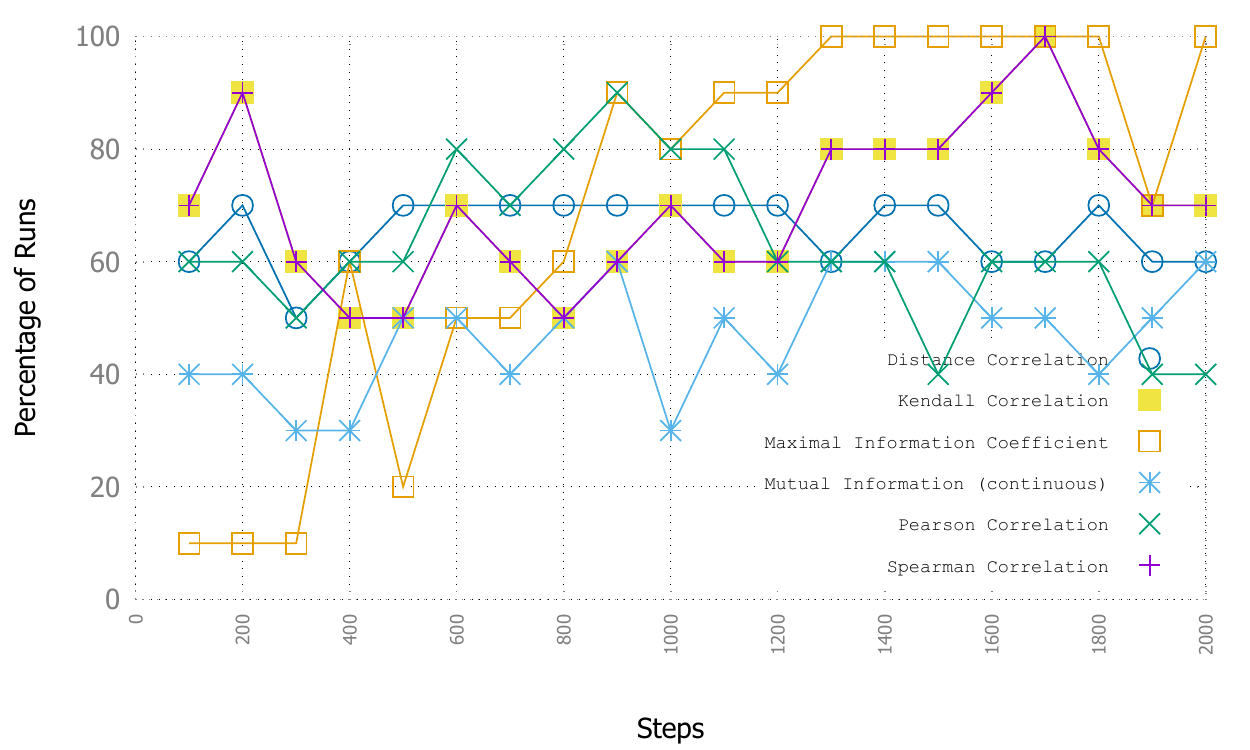}
		\caption{The results for the zoom.}
	\end{subfigure}
\caption[The results for scenario SCN 1 using the general method.]{The results for scenario \textit{SCN 1} using the general method. Each graph shows the fraction of runs in which the influence of the configuration component (pan, tilt, or zoom) of Camera 1 on Camera 0 is detected to be higher than the influence of Camera 2.}\label{fig:300-detection-of-influences_further-aspects_eval_camera-testDetection4-oneEstimator}
\end{figure}

The results are depicted in Figure~\ref{fig:300-detection-of-influences_further-aspects_eval_camera-testDetection4-oneEstimator}. There, we see that after a few 100 steps most measures allow a correct detection in $100\%$ of the cases. The continuous mutual information and maximal information coefficient on the other hand take a little longer to reach this level of certainty. For the pan, we see that the distance correlation shows the best result compared to the other measures. However, the other measures work as well with a higher sample size. Even though the MIC finds an influence quite reliably, the other measures do not show a definite result which can be explained by the minor influence of this configuration components, i.e., in most cases, the zoom does not determine if there is a positive reward at all but only the height of this positive reward.

Concluding the results, we have seen that it is possible to detect the influences in simple examples based on real-world applications, such as SC networks. We will see how to use this in scenarios with more complex requirements in the remainder of this article.\\

After the case that has been solved without the consideration of other configuration components, we focus on a scenario with few changes from scenario SCN 1. A top-down view on scenario SCN 2 is depicted in Figure \ref{fig:eval:camera:detectionStrong}. In the figure, we see three black dots that mark the position of the cameras. Camera 1 and Camera 2 are quite close and potentially share a common field of view which is marked by the colors areas around the cameras. Camera 3 on the other hand is separated and cannot observe the same areas as the other two cameras. The only difference to the previously evaluated scenario is where the objects of interest appear and move. In the last scenario, they were limited to the area between Camera 2 and 3 appearing from the south. Here, we see that they appear from the north and the south on the entire area around Camera 1 and 2, i.e., the two cameras are surrounded by objects. The influences from Camera 2 and Camera 3 on Camera 1 have been measured for the pan, the tilt, and the zoom. We expect that the pan and the tilt of Camera 2 have the most influence. Since a reward greater than 0 is much more likely in this scenario than in the previous, we also expect the zoom to play a more important role.

\begin{figure}	
	\centering
	\begin{subfigure}[t]{\textwidth}
		\centering
		\includegraphics[height=0.25\textheight]{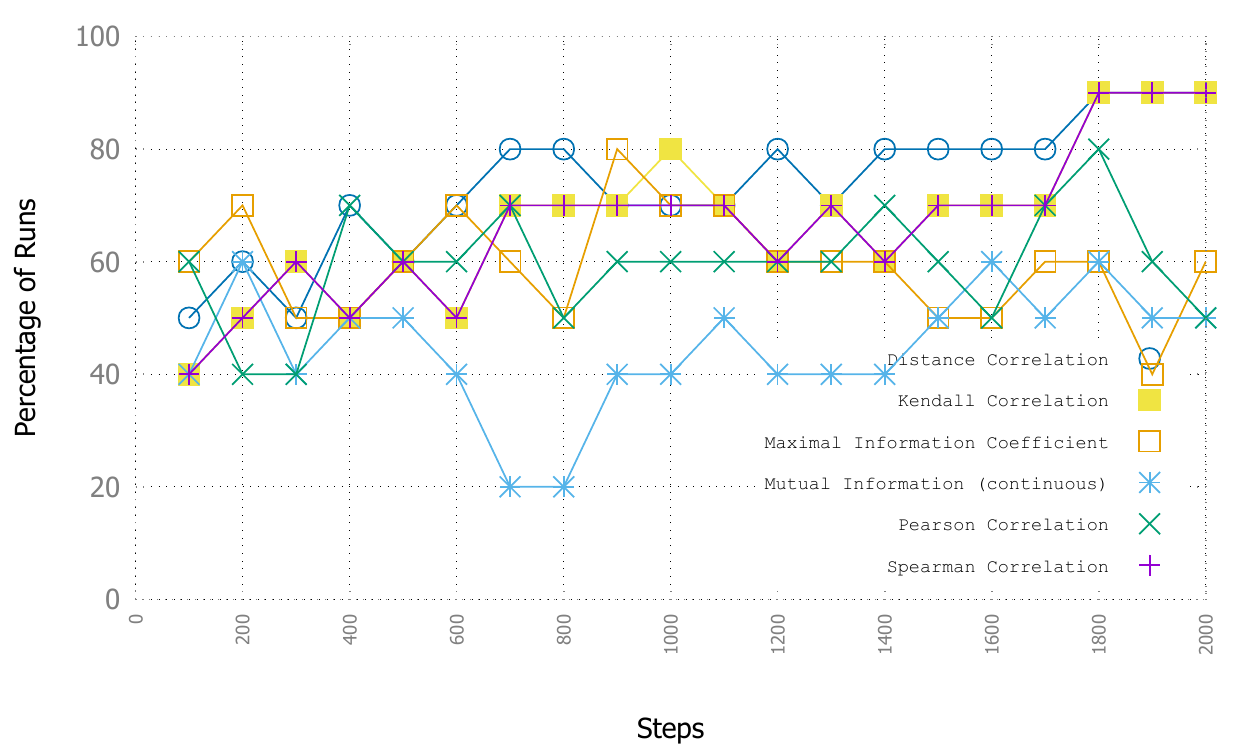}
		\caption{The results for the pan.}		
	\end{subfigure}
	\\
	\begin{subfigure}[t]{\textwidth}
		\centering
		\includegraphics[height=0.25\textheight]{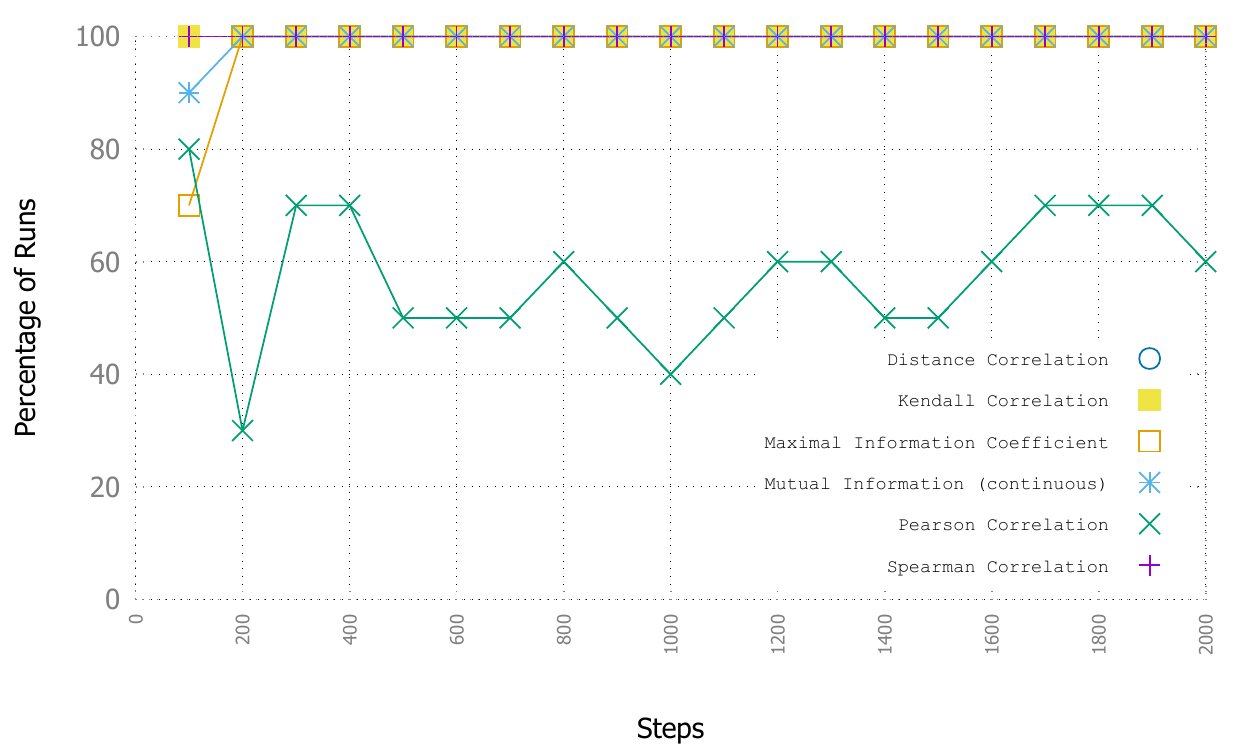}
		\caption{The results for the tilt.}
	\end{subfigure}
	\\
	\begin{subfigure}[t]{\textwidth}
		\centering
		\includegraphics[height=0.25\textheight]{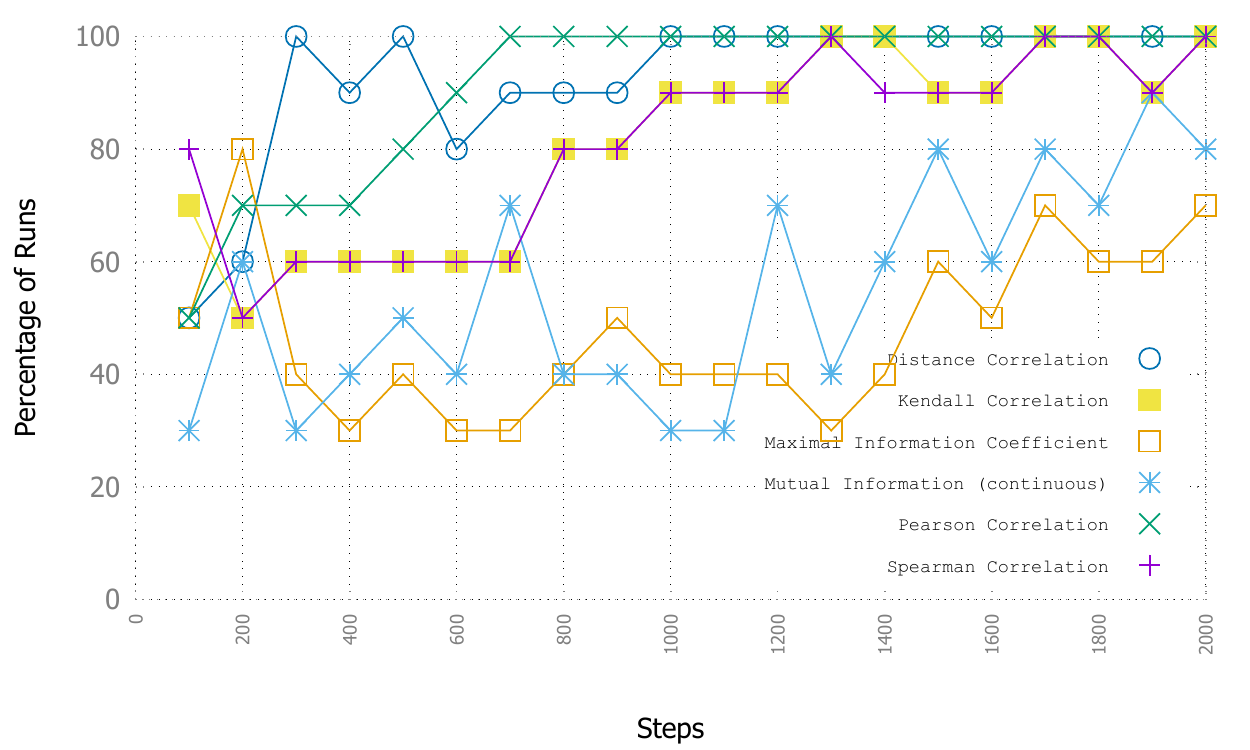}
		\caption{The results for the zoom.}
	\end{subfigure}
	\caption[The results for the scenario SC2 using no conditioning.]{The results for the scenario SC2 using no conditioning. Each graph shows the fraction of runs in which the influence of the configuration component (pan, tilt, or zoom) of Camera 1 on Camera 0 is detected to be higher than the influence of Camera 2.}\label{fig:300-detection-of-influences_further-aspects_eval_camera-detectionStrong:oneEstimator}
\end{figure}

\begin{figure}	
	\centering
	\begin{subfigure}[t]{\textwidth}
		\centering
		\includegraphics[height=0.25\textheight]{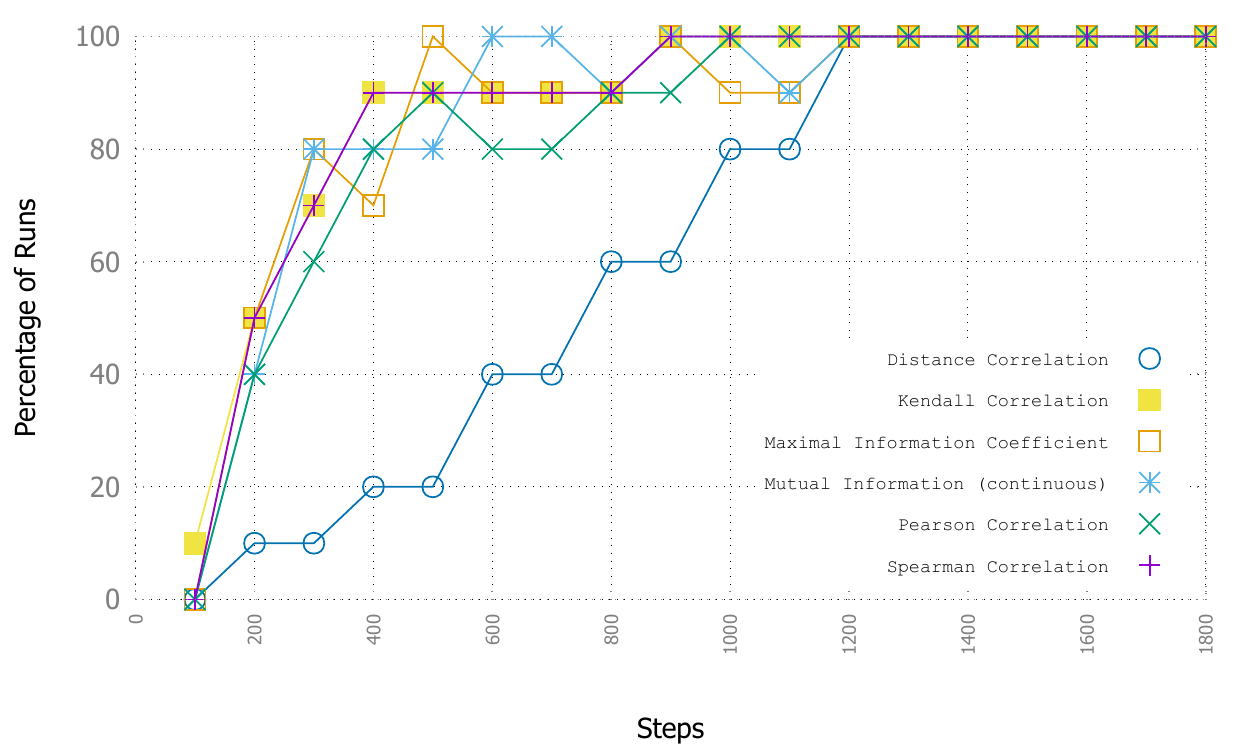}
		\caption{The results for the pan.}		
	\end{subfigure}
	\\
	\begin{subfigure}[t]{\textwidth}
		\centering
		\includegraphics[height=0.25\textheight]{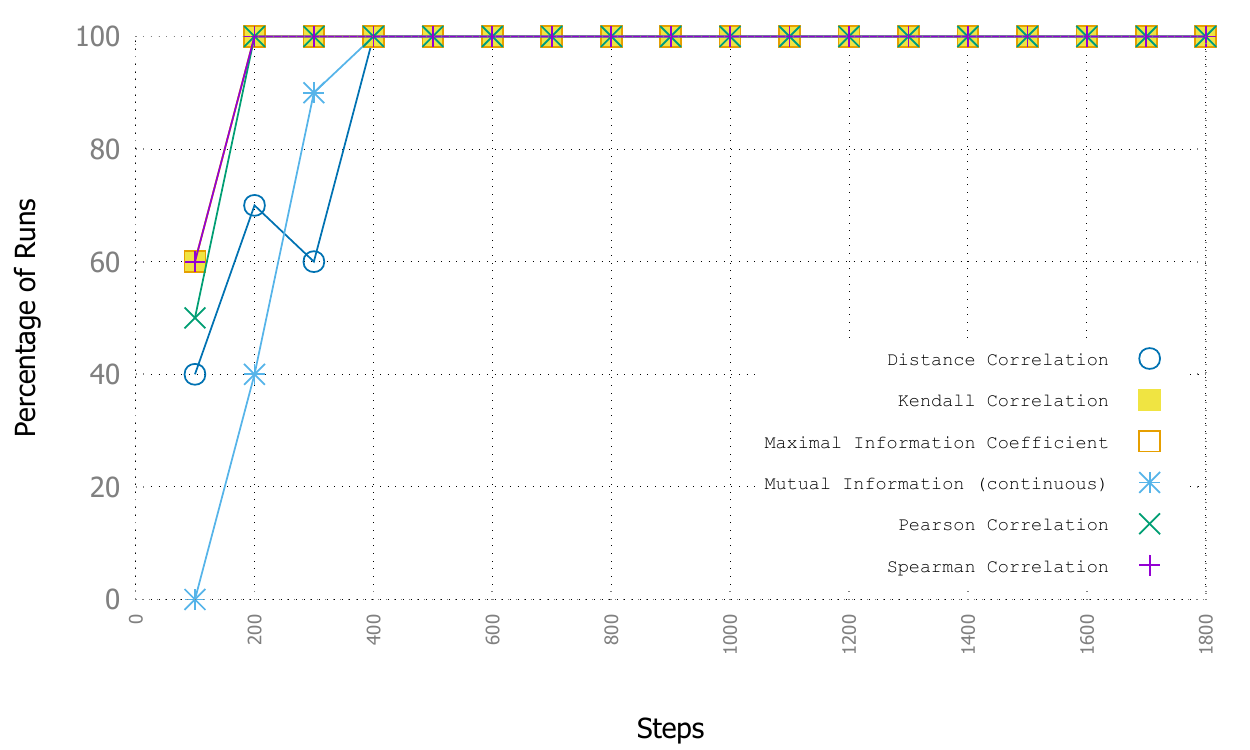}
		\caption{The results for the tilt.}
	\end{subfigure}
	\\
	\begin{subfigure}[t]{\textwidth}
		\centering
		\includegraphics[height=0.25\textheight]{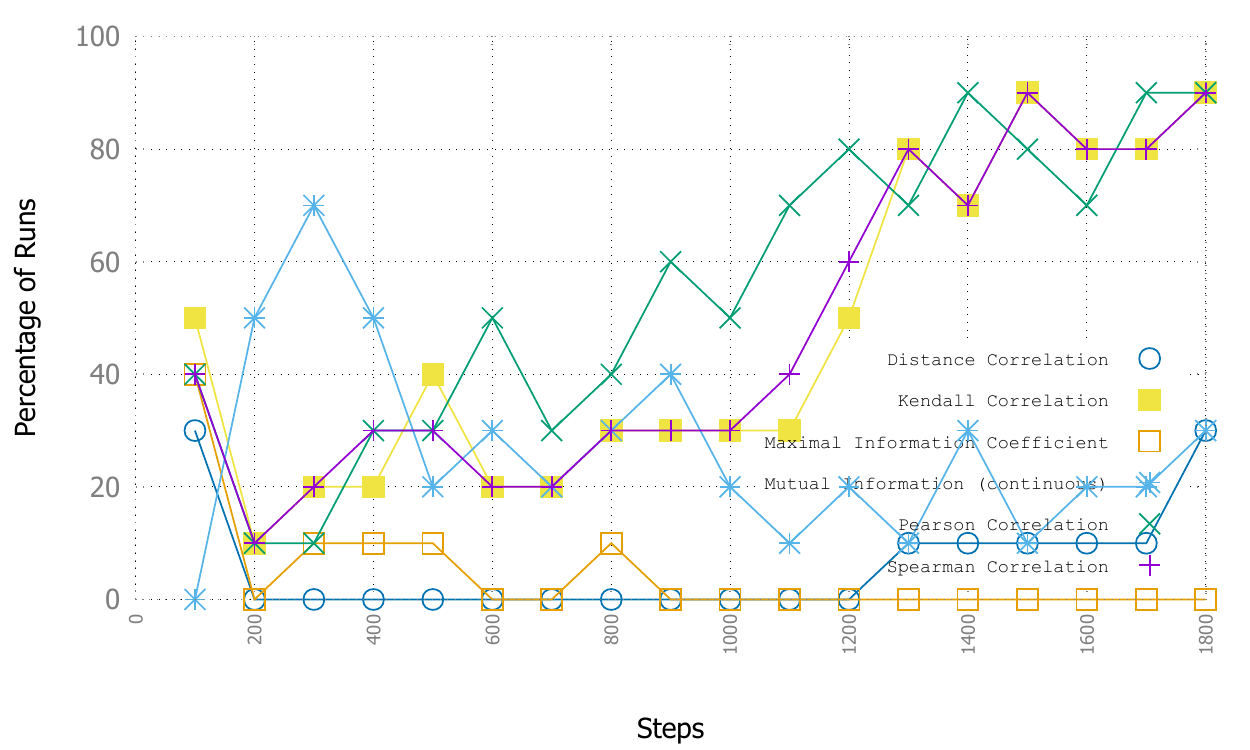}
		\caption{The results for the zoom.}
	\end{subfigure}
	\caption[The results for the scenario SC2 using two parts for conditioning.]{The results for the scenario SC2 using two parts for conditioning. Each graph shows the fraction of runs in which the influence of the configuration component (pan, tilt, or zoom) of Camera 1 on Camera 0 is detected to be higher than the influence of Camera 2.}\label{fig:300-detection-of-influences_further-aspects_eval_camera-detectionStrong:twoEstimator}
\end{figure}

The results are depicted in Figure \ref{fig:300-detection-of-influences_further-aspects_eval_camera-detectionStrong:oneEstimator} and \ref{fig:300-detection-of-influences_further-aspects_eval_camera-detectionStrong:twoEstimator}. As before, 10 independent runs have been conducted and we see the ratio of runs in which the configuration components of Camera 2 have been identified as more influencing than those of Camera 3. In Figure \ref{fig:300-detection-of-influences_further-aspects_eval_camera-detectionStrong:oneEstimator}, we see the results for the previous method. Even though the camera placement is identical to the previous scenario, we see that the pan is not detected within the first 2000 steps. Furthermore, the tilt cannot be detected by the Pearson correlation. This is because it is necessary to consider the other configuration components in this scenario. Therefore, we adopted the conditioned calculation, i.e., we split each of the other configurations in two parts and sort the points in different buckets depending on which configuration has been assumed. Since we face 5 other configurations, 
this leads to $2^5=32$ buckets. The values calculated for the buckets are then added up to get an aggregated result. The results achieved by this method are depicted in Figure \ref{fig:300-detection-of-influences_further-aspects_eval_camera-detectionStrong:twoEstimator}. We see that using this method the issues experienced in the first experiment do not appear and a flawless detection is ensured.

Concluding the results, we have seen that small changes in the setting can make it necessary to consider the other configuration components in the influence detection. A fast and reliable detection can be achieved by conditioning of the dependency measures.

\afterpage{\clearpage}
\afterpage{\clearpage}

%
%
\section[Influence Detection at Runtime]{Influence Detection at Runtime}\label{sec:300-detection-of-influences_further-aspects_runtime}
Modern ICT systems often cannot be tested entirely at design-time since not all circumstances the system faces can be foreseen due to a lack of awareness or time constraints. This applies especially to the type of systems that this article focuses on: self-adaptive and self-organising systems acting in shared environments. Obviously, it is hardly possible to foresee all situations a system will face if the considered system has to act in parallel to other systems (that may be under control of another stakeholder) and both should be able to adapt to new goals or environmental changes. Therefore, this section shifts the focus towards the detection of influences at runtime. A special examination of this is satisfied because of the problem of static behavior of the systems that introduces disturbances in the measurement by causing correlations that are not based on causality but coincidence. 

%
%
\subsection{Methodology for Detection}
Regarding the detection at runtime, we have to consider that the proposed influence detection algorithm relies on dependency measures that estimate the correlation between two random variables. If the method is applied at runtime, this can lead to wrongly detected influences because such correlations can appear without an underlying causality. {\color{black}The previous experiments demonstrated that it is possible to infer a causality from the correlation between the actions of two subsystems, since we enforced the configurations to be randomly selected from independent uniform distributions.} However, for the detection at runtime it has to be factored in that we can face autocorrelations and other disturbances within the configurations of the systems.

The approach to avoid this is to rely mostly on randomized configuration that appear naturally in applications that use reinforcement learning during runtime. This is due to the need of exploration in these tasks to find optimal strategies and to avoid to "get stuck" in a local optimum. This happen often with greedy approaches because the algorithm only tries one behavior and if it is ``good'' it sticks to it and misses out on other strategies that may lead to better results. There are different strategies to avoid this behavior. An easy and reliable approach is to use an $\epsilon$-greedy action selection, i.e., the algorithm will stick to the action that it has evaluated as the best one so far most of the time. However, with a probability of $\epsilon$ it will use a random action regardless of the so far evaluated usefulness of it. This means a natural approach to the issue of correlation without causality is to only use the samples that are formed from such exploration steps. But this would lead to a significant lower amount of samples. To avoid this we will also examine in detail how much ``randomness'' is necessary to allow the measurement to function properly by conducting experiments with different levels of $\epsilon$. Identifying the lowest level of $\epsilon$ will ensure the fastest detection of influences without running the risk of falsifying the measurement.

One might expect that a very high level is crucial for a successful detection. However, the correlation mainly appears due to the repetition of specific patterns in both systems in the same frequency. If such patterns are brought out of sync, a correct detection is possible despite the repetition, i.e., a high autocorrelation does not necessarily mean that the measurement is flawed. In example, two systems $A$ and $B$ each switch back and forth between their two configurations $1$ and $2$. The system $A$ gets a reward of $0.5$ if it applies $1$ and $0.8$ if it applies $2$. If both systems switch their configurations in the same time step the measurement will be falsified because system $B$'s configuration will correlate with the reward of system $A$. If one of the systems randomly does not switch the configuration the behavior becomes ``asynchronous'' and the measurement works well.

%
%
\subsection{Evaluation}
In the following, the evaluation of the runtime detection and the method to adapt to the influences are evaluated. \textcolor{black}{For the remainder, the evaluation parts of the article focus on the smart camera scenarios only. At this point, the consideration of the elementary use cases would not add to the discussion of the approach and are therefore omitted.}

We start on SCN 2 that has been introduced in Figure~\ref{fig:eval:camera:detectionStrong}. Briefly recapped, we have three cameras, where Camera 1 and Camera 2 influence each other due to an overlap in their potential observable space. Camera 3 does not overlap with one of the other cameras, and, therefore, it does not influence the other two. Previously, we saw how the detection works when the configuration is chosen randomly from the full configuration space.

In the following, the experiments are based on control mechanisms that use Q-learning, using discretized states and relative actions. Q-learning has been chosen since it is widely used and well understood. For the concrete implementation, we limit the alignment of the camera to $12$ pan angles, i.e., it is from $s_p=\{0,30,60,\dots,300,330\}$, to $3$ tilt angles, i.e., it is from $s_t=\{120,150,180\}$, and to $2$ zoom levels, i.e., it is from $c_z=\{12,18\}$. This results in $12\cdot 3\cdot 2 =72$ states for each camera. The camera can increase, decrease or leave the pan, tilt and zoom in each time step resulting in $3\cdot 3\cdot 3 = 27$ actions that can be applied. This means that the configuration space is $C=s_p\times s_t\times s_z$ for each camera.

\begin{figure}	
	\centering
	\begin{subfigure}[t]{0.7\textwidth}
		\centering
		\includegraphics[width=\textwidth]{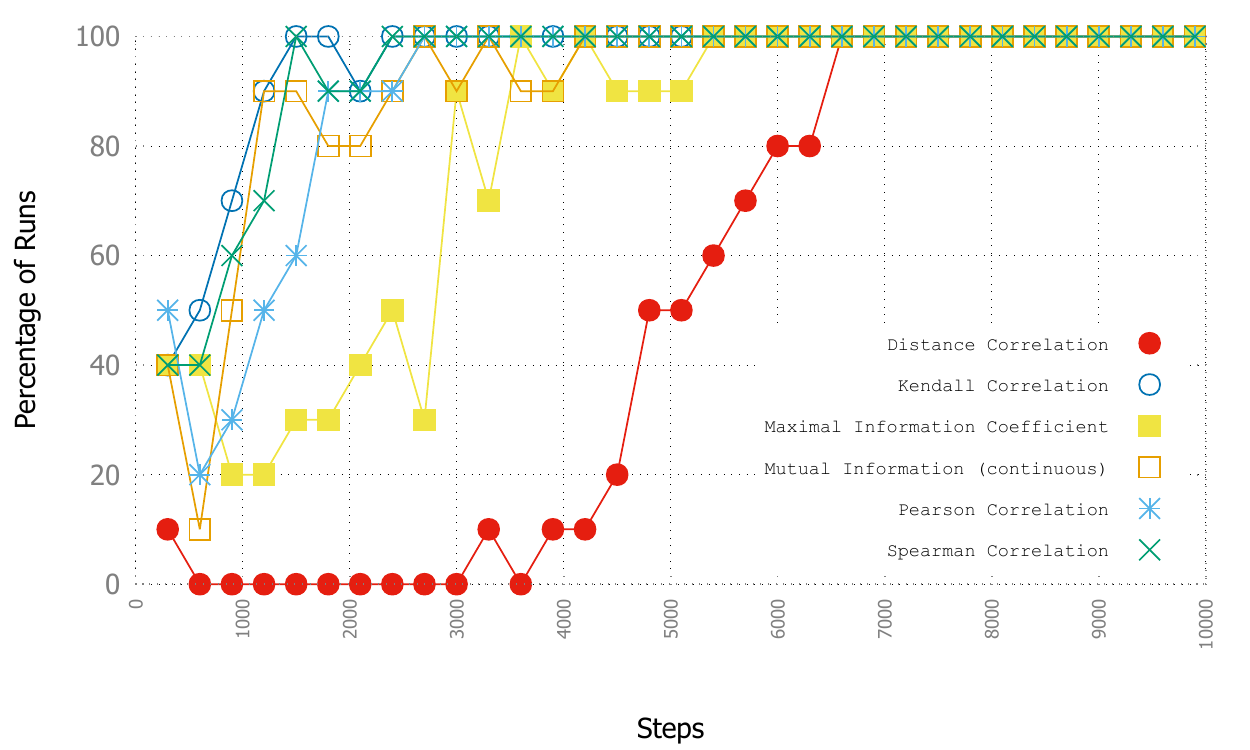}
		\caption{The results for the pan.}	
	\end{subfigure}
	\\
	\begin{subfigure}[t]{0.7\textwidth}
		\centering
		\includegraphics[width=\textwidth]{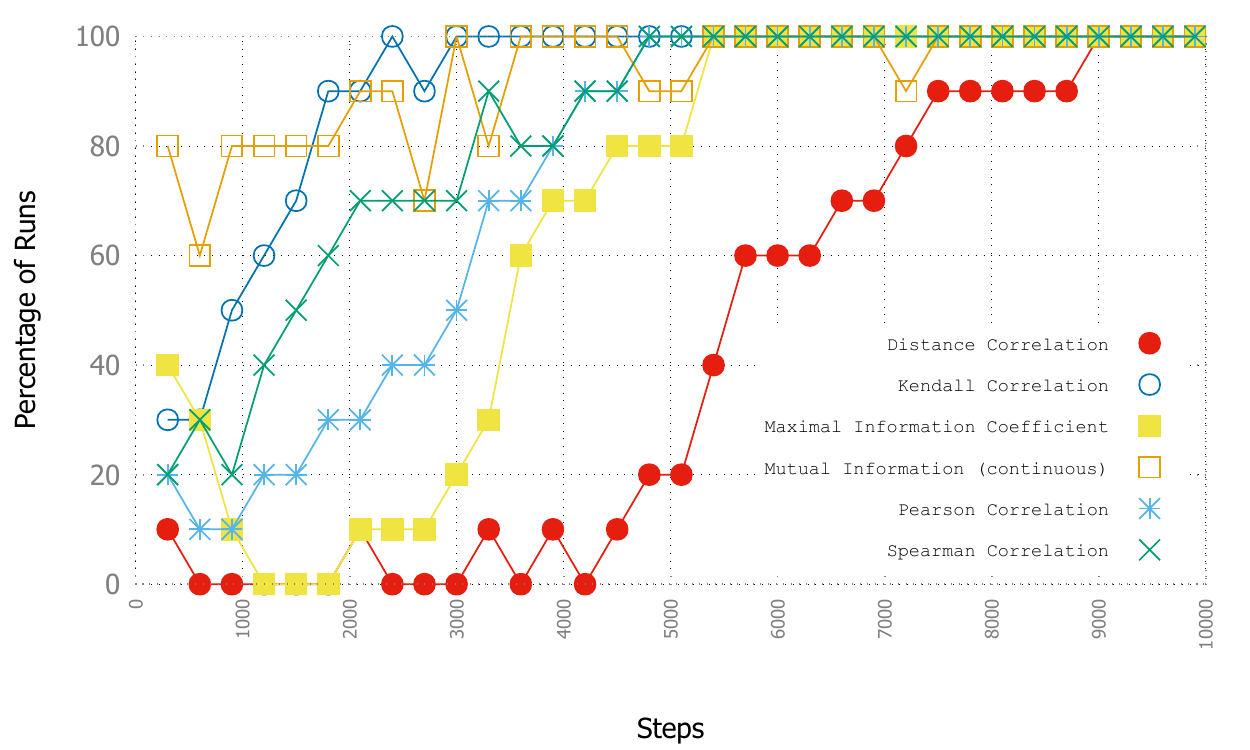}
		\caption{The results for the zoom.}
	\end{subfigure}
\caption[First part of the results for the detection of influences at runtime in scenario SCN 2 using a discrete state-action space with random actions applied.]{The results for the detection of influences in SCN 2. The influence detection rate for the \textbf{pan} and \textbf{zoom} of Camera 1 versus the Camera 2 based on $10$ independent runs. The measurement is similar to the graphs before but with the discretization of the state space and a Q-learning that applies the actions randomly. The graphs show the number of runs, in which the influence of Camera 1 is measured higher than the influence of Camera 2, for different numbers of samples.}\label{fig:300-detection-of-influences_further-aspects_adaption_detection-strong_baseline_pan-zoom}
\end{figure}

\begin{figure}	
	\centering
	\begin{subfigure}[t]{0.7\textwidth}
		\centering
		\includegraphics[width=\textwidth]{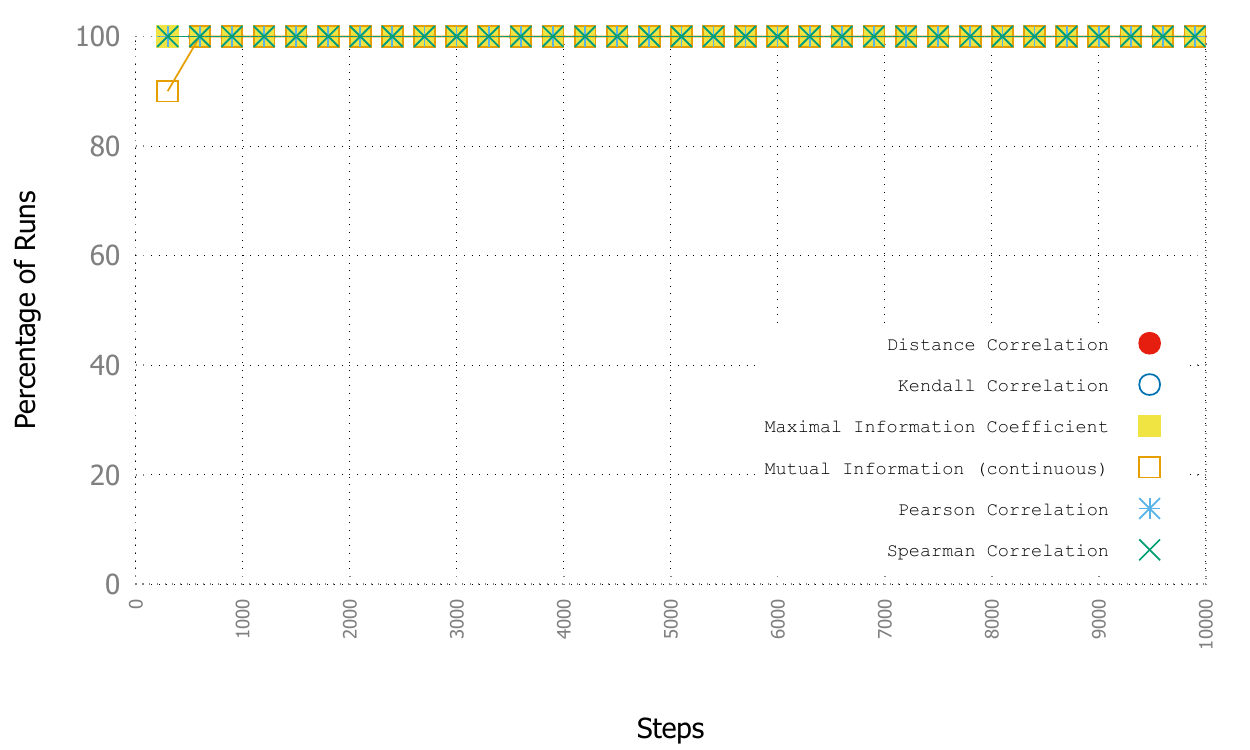}
		\caption{The results for $300$-$9900$ samples.}		
	\end{subfigure}
	\\
	\begin{subfigure}[t]{0.7\textwidth}
		\centering
		\includegraphics[width=\textwidth]{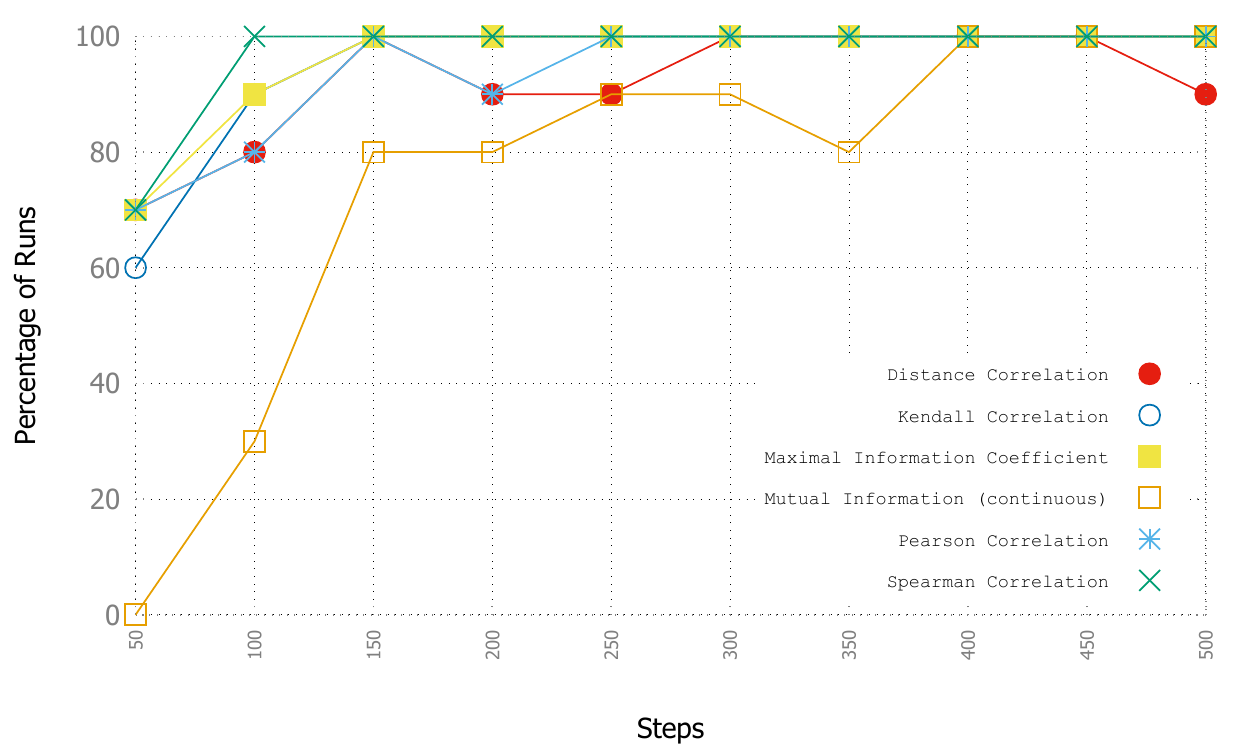}
		\caption{A detailed few for $50$-$500$ samples.}
	\end{subfigure}
\caption[Second part of the results for the detection of influences at runtime in scenario SCN 2 using a discrete state-action space with random actions applied.]{The results for the detection of influences in SCN 2. The influence detection rate for the \textbf{tilt} of Camera 1 versus the Camera 2 based on $10$ independent runs. The measurement is similar to the graphs before but with the discretization of the state space and a Q-learning that applies the actions randomly. The graphs show the number of runs, in which the influence of Camera 1 is measured higher than the influence of Camera 2, for different numbers of samples.}\label{fig:300-detection-of-influences_further-aspects_adaption_detection-strong_baseline_tilt}
\end{figure}

The first evaluation step is to find out how the limitation of the configuration space and the order of states that are visited affect the influence detection. The results can be found in Figure~\ref{fig:300-detection-of-influences_further-aspects_adaption_detection-strong_baseline_pan-zoom} and~\ref{fig:300-detection-of-influences_further-aspects_adaption_detection-strong_baseline_tilt}. The system has run for $10,000$ steps by applying random actions and the figure shows the percentage of runs in which Camera 1 detects Camera 2 as more influencing than Camera 3. The results for the pan and zoom are depicted in Figure~\ref{fig:300-detection-of-influences_further-aspects_adaption_detection-strong_baseline_pan-zoom}. There, we see that each measure can find the influences very reliably within the $10,000$ steps with the distance correlation being much more unreliable using less samples. In Figure~\ref{fig:300-detection-of-influences_further-aspects_adaption_detection-strong_baseline_tilt}, the results for the tilt are depicted. We can see that only few steps are needed to find the influence, i.e., in only 600 steps each of the measures can detect very reliable the influence.

\begin{table}[]
	\centering
	
    \begin{tabular}{|c|c|c|c|c|c|c|c|c|}
	    \hline
    	Step & Start & $10k$ & $20k$ & $30k$ & $40k$ & $50k$ & $70k$ & $80k$ \\
        \hline
		$\epsilon$ & $1.0$ & $0.9$ & $0.81$ & $0.73$ & $0.66$ & $0.59$ & $0.53$ & $0.48$ \\
        \hline
        \hline
        Step & $90k$ & $100k$ & $110k$ & $120k$ & $130k$ & $140k$ & $150k$ & $160k$ \\
        \hline
        $\epsilon$ & $0.43$ & $0.39$ & $0.35$ & $0.31$ & $0.28$ & $0.25$& $0.23$ & $0.20$  \\
        \hline
        \hline
        Step & $170k$ & $180k$ & $190k$ & $200k$ & $210k$ & $220k$ & $230k$ & $240k$ \\
        \hline
        $\epsilon$ & $0.19$& $0.17$& $0.15$ & $0.14$ & $0.12$ & $0.11$ & $0.1$ & $0.09$ \\
        \hline
        \hline
        Step & $250k$ & $260k$ & $270k$ & $280k$ & $290k$ & $300k$ & $310k$ & $320k$\\
        \hline
        $\epsilon$ & $0.08$ & $0.07$ & $0.06$ & $0.06$ & $0.05$ & $0.05$ & $0.05$ & $0.05$ \\
        \hline
    \end{tabular}
    \caption{The values for $\epsilon$ in the $\epsilon$-greedy action selection. It starts at $1.0$ and decreases by $10\%$ every $10k$ steps until it reaches $5\%$.}
    \label{tab:300-detection-of-influences_further-aspects_adaption_epsilon}
\end{table}
\afterpage{\clearpage}

\begin{figure}	
	\centering
	\begin{subfigure}[t]{\textwidth}
		\centering
		\includegraphics[height=0.25\textheight]{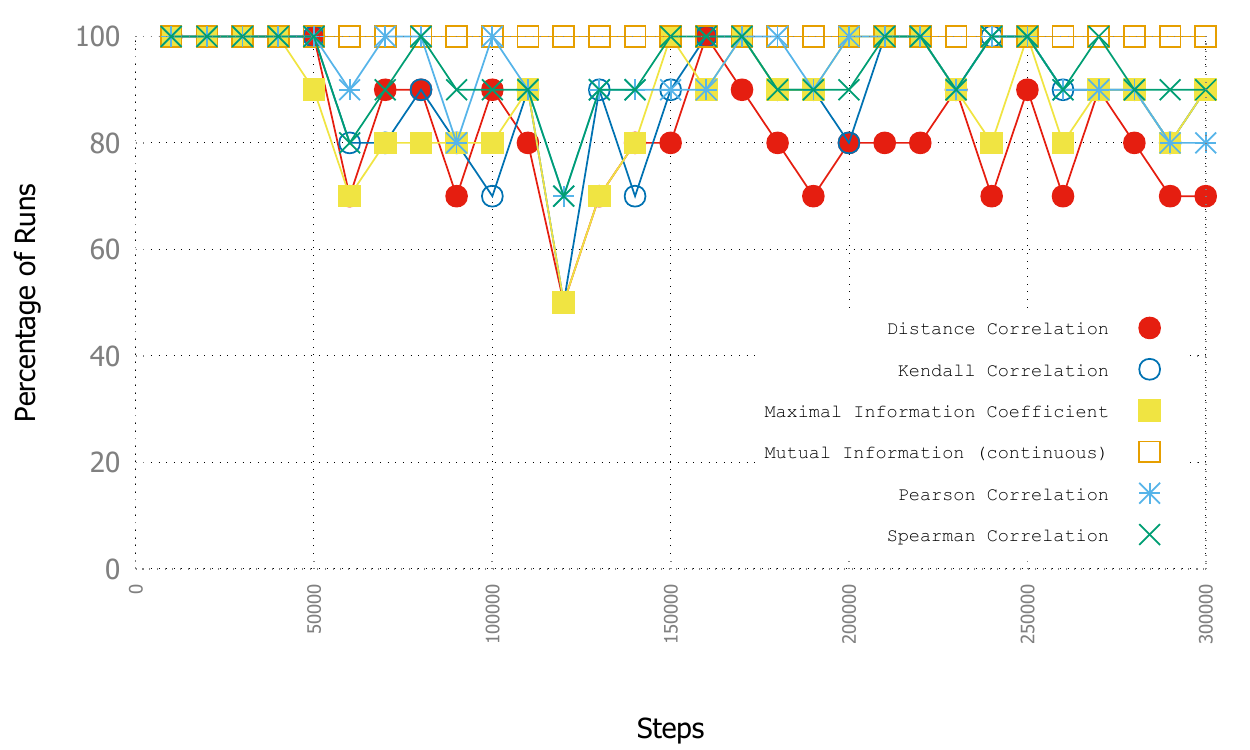}
		\caption{The results for the pan.}
		\label{fig:300-detection-of-influences_further-aspects_adaption_detection-strong_detection-for-different-epsilon_no-expand_pan}	
	\end{subfigure}
	\\
	\begin{subfigure}[t]{\textwidth}
		\centering
		\includegraphics[height=0.25\textheight]{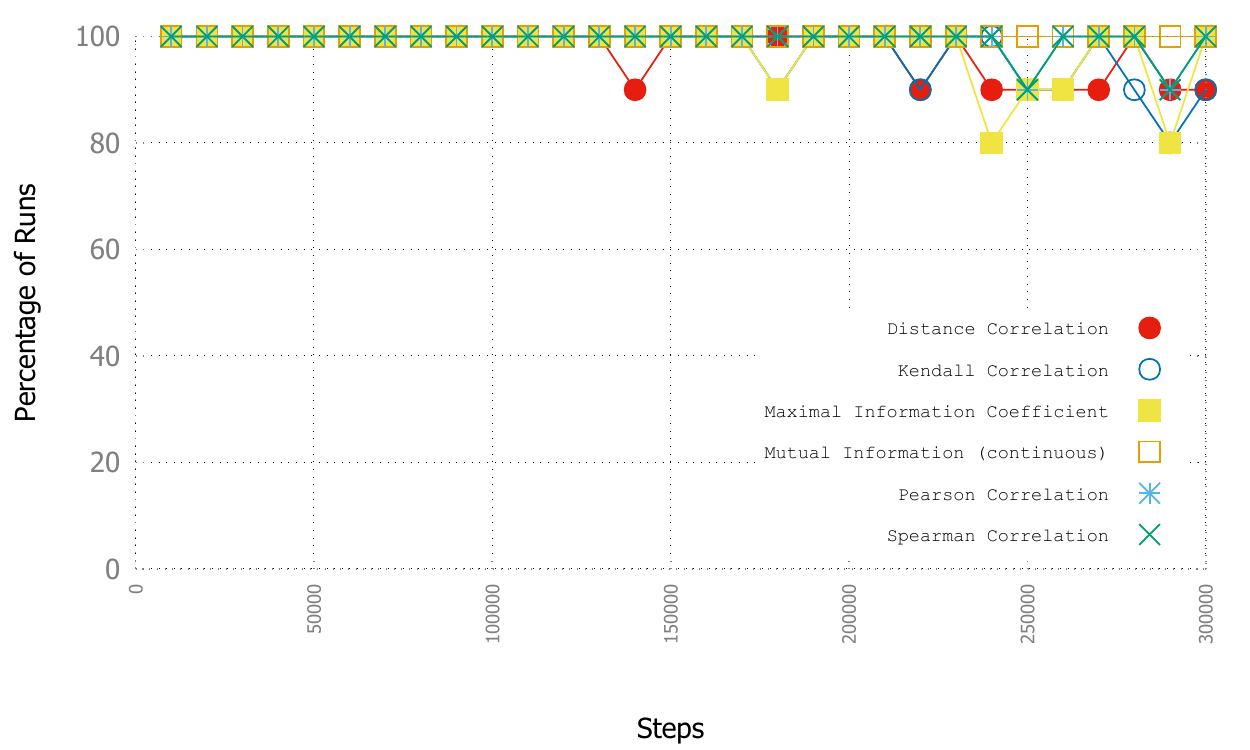}
		\caption{The results for the tilt.}
		\label{fig:300-detection-of-influences_further-aspects_adaption_detection-strong_detection-for-different-epsilon_no-expand_tilt}
	\end{subfigure}
	\\
	\begin{subfigure}[t]{\textwidth}
		\centering
		\includegraphics[height=0.25\textheight]{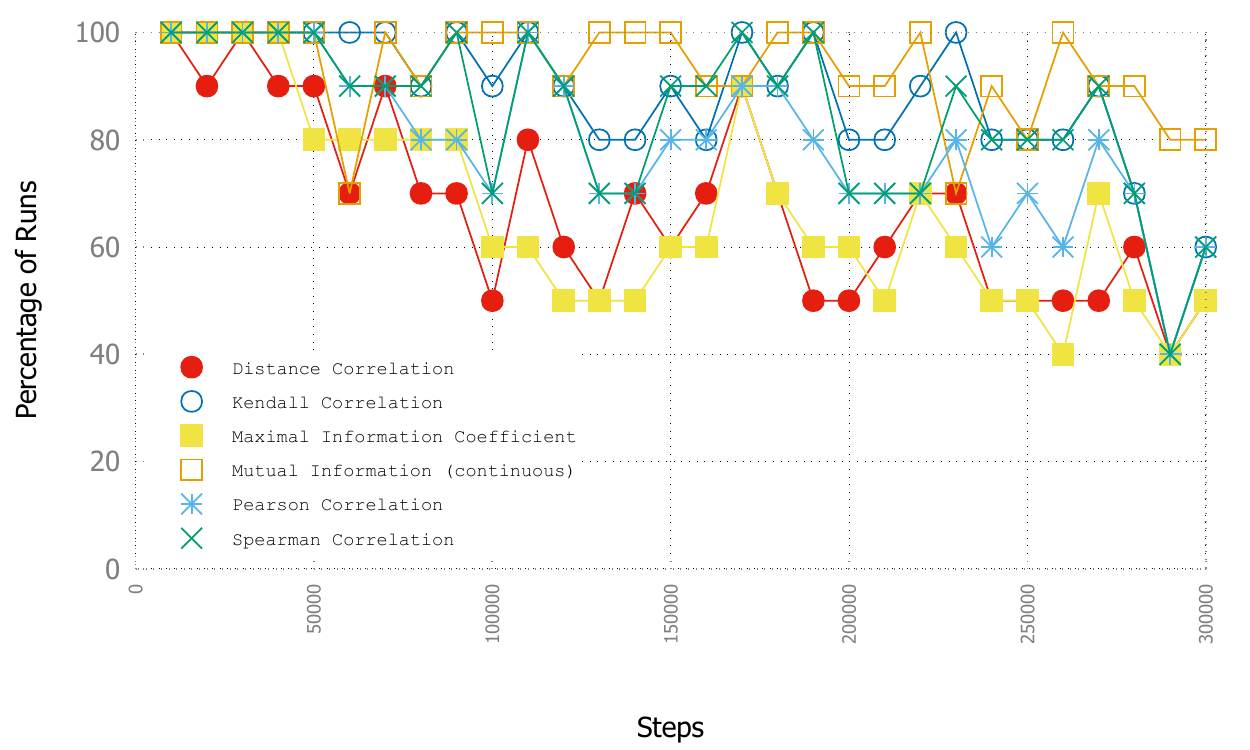}
		\caption{The results for the zoom.}
		\label{fig:300-detection-of-influences_further-aspects_adaption_detection-strong_detection-for-different-epsilon_no-expand_zoom}
	\end{subfigure}
\caption[First part of the results for the detection of influences at runtime in scenario SCN 2 using a discrete state-action space with $\epsilon$-greedy strategy.]{The results for the detection of influences in SCN 2. The influence detection rate for the pan, tilt and zoom of Camera 1 versus the Camera 2 based on $10$ independent runs. The measurement is similar to the graphs before but here the camera is controlled by a Q-learning algorithm with $\epsilon$-greedy strategy, where the $\epsilon$ is falling from 1 to 0.05 in 10\%-declines each $10k$ steps. The graphs show the number of runs, in which the influence of Camera 1 is measured higher than the influence of Camera 2, using the last $10k$ steps as samples.}\label{fig:300-detection-of-influences_further-aspects_adaption_detection-strong_detection-for-different-epsilon_no-expand_pan-tilt-zoom}
\end{figure}

In the next step, we analyze how the detection rate changes if we switch from a purely random action selection to $\epsilon$-greedy. The $\epsilon$ is varied between $1$ and $0.05$ according to Table~\ref{tab:300-detection-of-influences_further-aspects_adaption_epsilon}. It reflects that $\epsilon$ is set to $1$ at start and is then decreased by $10\%$ each $10,000$ steps. The results can be found in Figure~\ref{fig:300-detection-of-influences_further-aspects_adaption_detection-strong_detection-for-different-epsilon_no-expand_pan-tilt-zoom}. There, we see the detection rate if single-agent Q-learning with an $\epsilon$-greedy action selection is used and the last $10,000$ samples are used for the detection. The Q-learning algorithm uses a low $\alpha=0.1$ to handle fluctuations in the reward signal and a high $\gamma=0.9$ to allow the learning of a sequence of actions. Figure~\ref{fig:300-detection-of-influences_further-aspects_adaption_detection-strong_detection-for-different-epsilon_no-expand_pan} illustrates that for the pan the detection works well for the first $50,000$ steps, i.e., for $\epsilon>0.66$. Afterwards, the detection works worse for some of the measures with an average of about $80$\%. The continuous mutual information sticks out here with a perfect detection even with low values of $\epsilon$. For the tilt, in Figure~\ref{fig:300-detection-of-influences_further-aspects_adaption_detection-strong_detection-for-different-epsilon_no-expand_tilt}, we see a similar effect, but it is way less distinct and occurs only for lower $\epsilon$. Figure~\ref{fig:300-detection-of-influences_further-aspects_adaption_detection-strong_detection-for-different-epsilon_no-expand_zoom} shows the graph for the zoom, which turns out to be way more sensitive regarding $\epsilon$ and makes the measurement unreliable.

Concluding the results, we have seen how the application of a control algorithm affects the influence detection in contrast to a random selection of configurations. Due to the disturbances and autocorrelations, it is possible that influences are not detected correctly. However, if it is ensured that an adequate amount of ``randomness'' is used to select the actions it results in proper results. For the example examined here, we have seen that the usage of 66\% randomly selected actions allows for a precise identification of influencing systems. If the usage of less randomly selected actions is necessary, an amount of greedy actions should be removed from the set for the calculation of influences until the required amount of "randomness" can be ensured.

\afterpage{\clearpage}

\afterpage{\clearpage}

\afterpage{\clearpage}

%
%
\section{Self-adapting to Influences}\label{sec:300-detection-of-influences_further-aspects_adaption}
In the previous sections, we focused on the detection of the influences, which is a crucial step towards the goal of optimal behavior. In this section, we introduce an approach to self-adaption to these identified influences, i.e., the exploitation of the influences if they have been discovered. For this adaption, the focus is not on the presentation of a single algorithm, but on giving a general methodology based on the RL model and realize it with Q-learning.

%
%
\subsection{Methodology for Self-adaption}

\begin{algorithm}[t]
 initialization\;
 \For{each step}{
  observe own configuration and reward\;
  distribute observations to other subsystems\;
  \If{enough observations}{
   estimate influences\;
   evaluate estimated influences\;
   \If{influence found}{
    adapt learning algorithm\;
   }
  }
  learning algorithm decides next action\;
  update learning algorithm with new observation\;
 }
 \caption{The main loop for each subsystem using the influence detection and adaption at runtime. 
 }\label{algo}
\end{algorithm}

In general, there is a variety of options available to address the influences since the detection is designed to work independently from the control algorithm of the system. Possible reactions to influences range from a hand-crafted solution during design-time to a system that adapts during runtime in static patterns and finally to a full self-learning behavior. \textcolor{black}{For the latter, we get an impression on how the subsystem's learning algorithm and influence detection workflow (cf. Figure \ref{fig:influence-detection:workflow}) interact by studying Listing \ref{algo}.} Potential candidates for such a learning algorithm can be found in the area of MARL. However, we rely on a basic principle that unifies three properties: 
\begin{enumerate}
\item It is possible to start out with single learners and later on learn the cooperation,
\item it can be applied to many reinforcement learning algorithms,
\item it can be used with systems that have heterogeneous configuration spaces and control algorithms, e.g., some of the systems have a static behavior.
\end{enumerate}
\textcolor{black}{
The basic adaption principle is depicted in Figure~\ref{fig:adaption} and shows how the \textit{adapt learning algorithm} from Listing \ref{algo} has been realized.
} 
There, we adopt the basic RL model introduced in Section~\ref{sec:200-system-model_relations_to_RL}. We assume that previous measurements have resulted in a detection of an influence from system $A$ on system $B$ meaning $B$ should react properly regarding the configuration of $A$. Therefore, we integrate the value of the configuration $c_t$ at time $t$ into the state $s_t$ resulting in a state $s'_t$. For example, if $B$'s states are given through a position represented as $(x,y)$-coordinate between 0 and 1 each, its state space is $S=\left[0,1\right]\times\left[0,1\right]$. Assuming the influence detection has found $A$'s configuration component $c_B$, which represents the speed of B and is between $-1$ and $1$, as influencing, the state space $S$ will be extended to a new state space $S'=S\times\left[-1,1\right]=\left[0,1\right]\times\left[0,1\right]\times\left[-1,1\right]$.

{
	\tikzstyle{agent} = [rectangle, rounded corners, ultra thick,  minimum width=3cm, minimum height=1cm,text centered, draw=black, fill=ACMOrange!30]
	\tikzstyle{environment} = [rectangle, rounded corners, ultra thick,  minimum width=3cm, minimum height=1cm,text centered, draw=black, fill=ACMLightBlue!30]
	\tikzstyle{arrowRL} = [ultra thick,->,>=stealth]	
	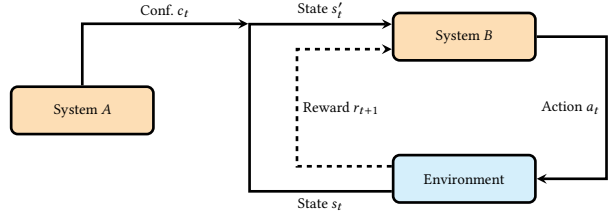
\begin{figure}[t]
		\centering
		\resizebox{8cm}{!}{%
		\begin{tikzpicture}[node distance=3cm]
		
		\node (agentB) [agent] {System $B$};
		\node (environment) [environment, below of=agentB] {Environment};
		
		\draw [arrowRL] (agentB) --++(+3,0) |- node [pos=0.25,left] {Action $a_t$} (environment);
		
		\path (environment.west) -- (environment.north west) coordinate[pos=0.5] (environmentHelperTop);
		\path (agentB.west) -- (agentB.south west) coordinate[pos=0.5] (agentHelperBottom);
		\draw [arrowRL,dashed] (environmentHelperTop) --++ (-2,0) |- node [pos=0.25,right] {Reward $r_{t+1}$} (agentHelperBottom);
		
		\path (environment.west) -- (environment.south west) coordinate[pos=0.5] (environmentHelperBottom);
		\path (agentB.west) -- (agentB.north west) coordinate[pos=0.5] (agentHelperTop);
		\draw [arrowRL] (environmentHelperBottom) --++ (-3,0) node [pos=0.5,below] {State $s_t$} |- node [pos=0.75,above] {State $s'_t$} (agentHelperTop);

		\node (agentA) [agent, below left =0.5cm and 5.0cm of agentB] {System $A$};
		\path (agentA.east) -- (agentA.north east) coordinate[pos=0.5] (agentAHelper);
		\draw [arrowRL] (agentA.north) --++ (0,+1.3) --++ (+3.5,0) node [pos=0.5,above] {Conf. $c_t$};
		
		\end{tikzpicture}
		}
		\caption{The proposed method to adapt to other influencing agents in terms of the reinforcement learning model.}
		\label{fig:adaption}
	\end{figure}
}

There are two points to discuss in this approach. The first one is how the influenced system $B$ will know the current configuration of $A$. This is not an issue with delayed influences but only with immediate. It can be resolved by letting the influenced system wait until it can observe the decision either by a message of the influencing system or through sensors. If several influences in the system are detected this can lead to chains of systems that wait on each other and it is necessary to check for each of the integrations in the state space that the graph they form is acyclic.

As mentioned before, this method is independent from the learning algorithm as long as it fits the RL model. But the extension of the state space can be seen as a special instance of transfer learning in the RL domain~\cite{TaylorS2009}. This makes algorithms that have a strong transfer capability more suitable for this task. In this article, we rely on Q-learning since it gives an easy and natural way to realize the transfer during runtime that will be explained in the following: Assuming that $c_B$ can take $k$ different values from each state $s$, there will be $k$ new states $s'_i$, $i\in\{1,\dots,k\}$. Before the expansion of the state space, the system holds a current Q-value for $Q(s,a_j)$, where $a_j\in A$ are the possible actions. To transfer the already learned knowledge, the new Q-values are set to $Q(s'_i,a_j):=Q(s,a_j)$, for each $i\in\{1,\dots,k\}$ and $a_j\in A$. The advantage here is that in states that are not affected by the other system the Q-value will remain at the correct value. In turn, in states where the system is influenced the values will be updated to their true value using the upcoming experiences.

If the state space is continuous the standard Q-learning can not be used. An alternative would be for example the extended classifier system (XCS-R) with continuous states~\cite{Wilson1995,Wilson2000} that has good online learning capabilities and is transferable~\cite{LiY2016}.

A further issue is that in the previous experiments about runtime detection, we have seen that a small amount of runs might detect the independent systems as more influencing than they actual are. This behavior is temporary and can be corrected by using a bigger sample size. However, during the runtime of the system, it is not possible to determine if it is a malicious detection. Therefore, we introduce a factor by that the real measurement has to be higher than the calculated notional independent counterpart which is used as baseline. Furthermore, the different configuration parts are ranked and after each influence calculation only the highest ranked configuration is integrated in the state space of each camera. \textcolor{black}{Even though this mechanism is likely to catch most erroneous integration, they cannot be avoided entirely. But, if a wrong integration is detected later, it can be corrected easily. To revert the false state space extension the new created states are merged back together by averaging the Q-values of the states, i.e., $Q(s,a_j)=\sum_{i=1}^k Q(s'_i,a_j)$ for every action $a_j\in A$.}

%
%
\subsection{Evaluation}
\begin{figure}	
	\centering
	\begin{subfigure}[t]{\textwidth}
		\centering
		\includegraphics[height=0.25\textheight]{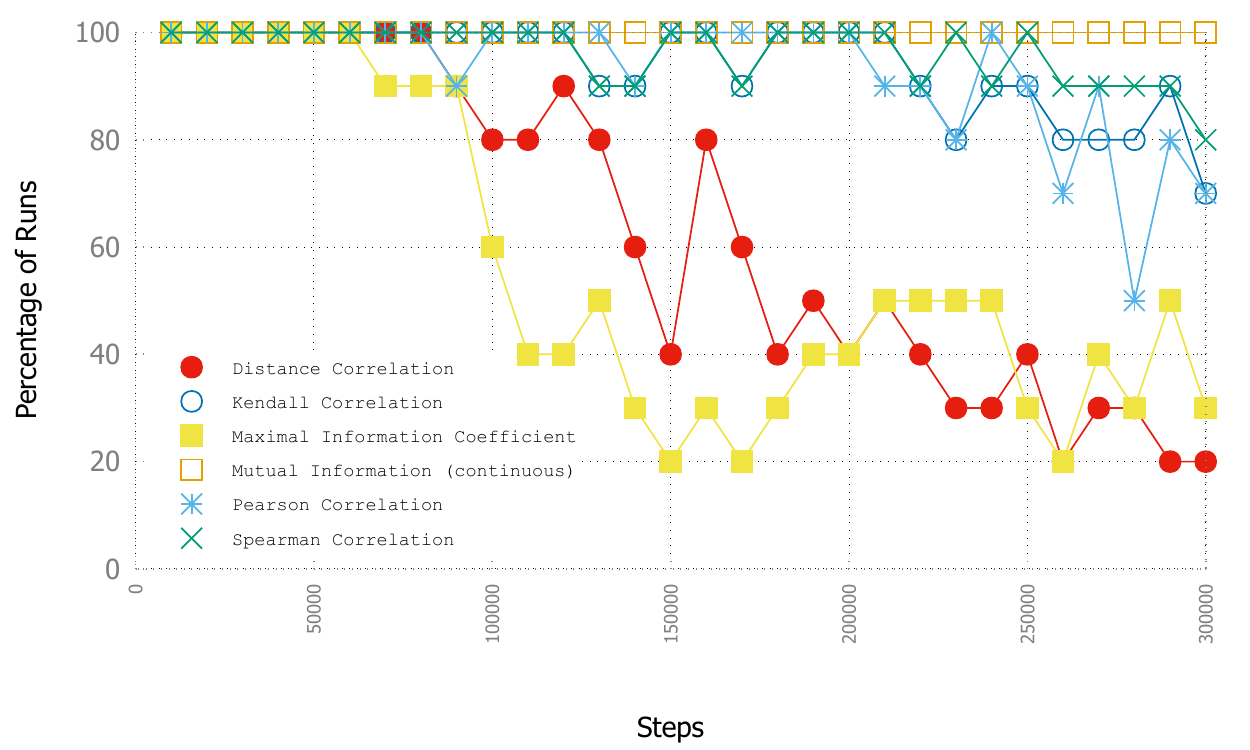}
		\caption{The results for the pan.}		
	\end{subfigure}
	\\
	\begin{subfigure}[t]{\textwidth}
		\centering
		\includegraphics[height=0.25\textheight]{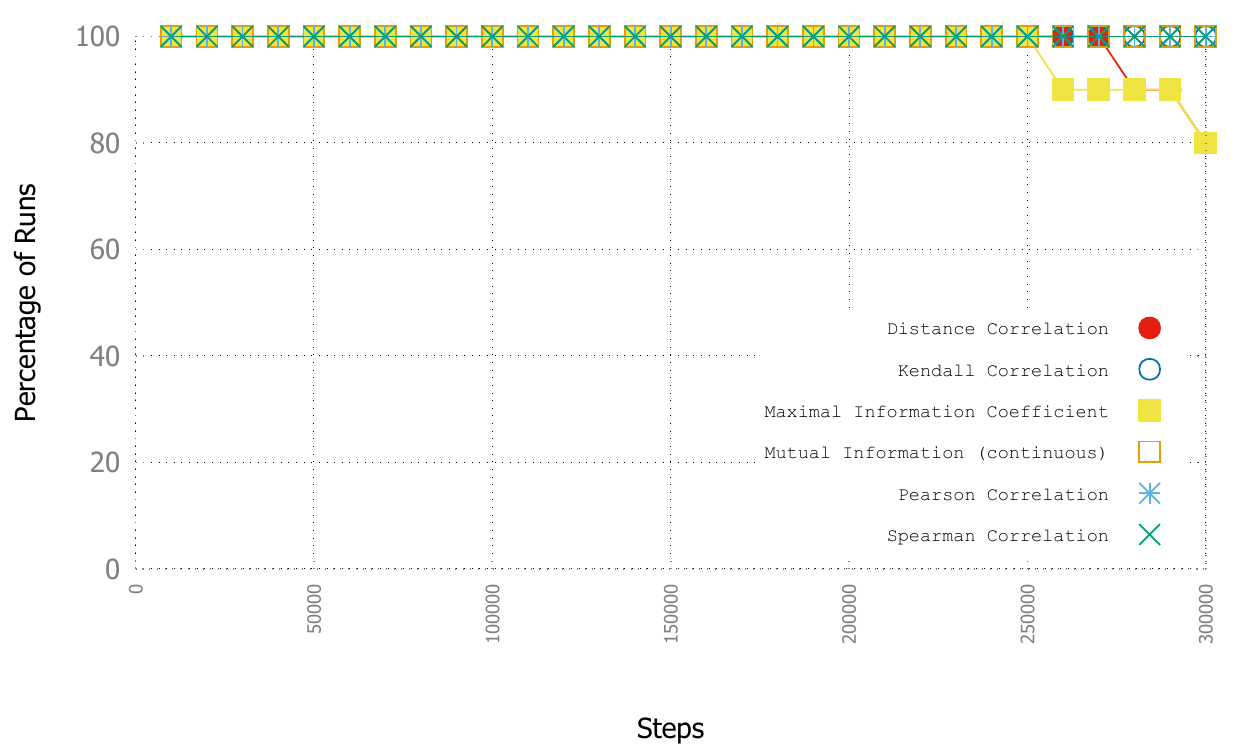}
		\caption{The results for the tilt.}
	\end{subfigure}
	\\
	\begin{subfigure}[t]{\textwidth}
		\centering
		\includegraphics[height=0.25\textheight]{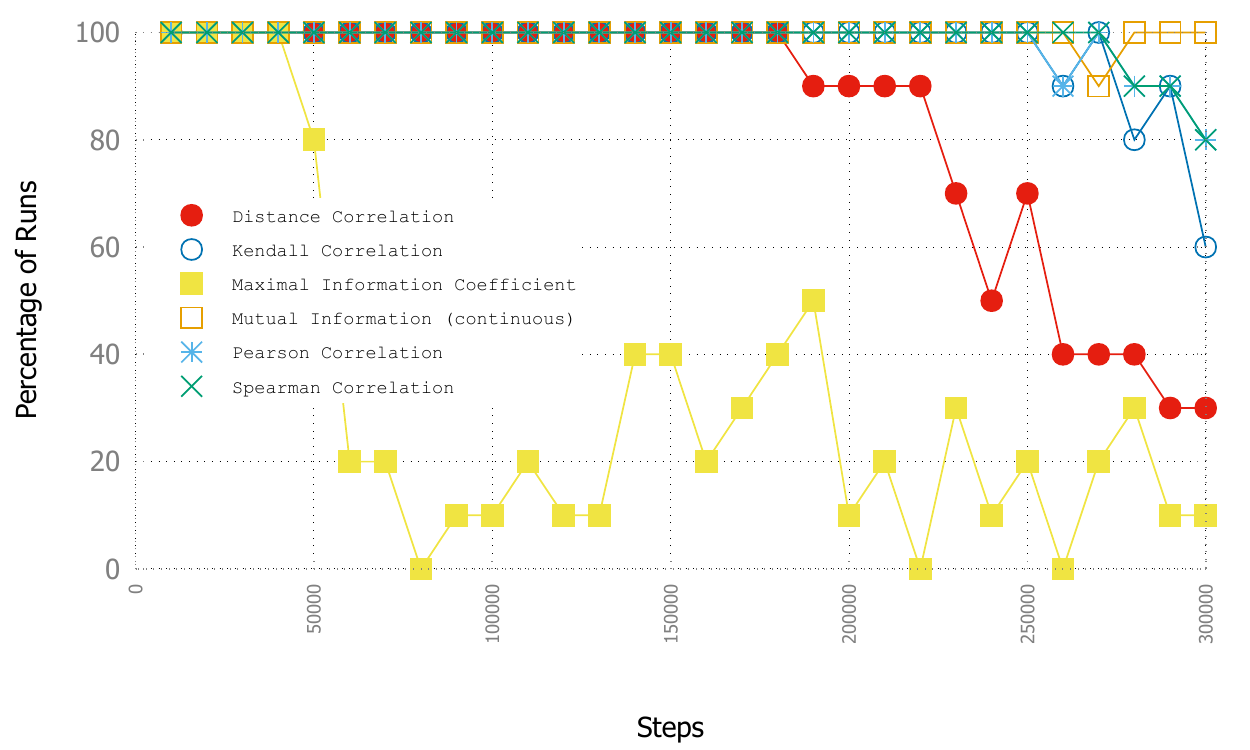}
		\caption{The results for the zoom.}
	\end{subfigure}
\caption[First part of the results for the detection of influences in SCN 2 at runtime with adaption to influences at design time.]{The results for the detection of influences in SCN 2. The influence detection rate for the pan, tilt and zoom of Camera 1 versus the Camera 2 based on $10$ independent runs. The cameras are controlled with the same algorithms as before but the pan, tilt, and zoom configuration of Camera 2 are integrated in the state space of Camera 1 at the start of the runs. The graphs show the number of runs, in which the influence of Camera 1 is measured higher than the influence of Camera 2, using the last $10k$ steps as samples.}\label{fig:300-detection-of-influences_further-aspects_adaption_detection-strong_detection-for-different-epsilon_expand-at-start_pan-tilt-zoom}
\end{figure}

In the next part, we examine how the influence measurement reacts to the adaption that has been introduced previously. To do so, we integrated the three configuration components of Camera 2 in the state of Camera 1 at the start of the simulations. In Figure~\ref{fig:300-detection-of-influences_further-aspects_adaption_detection-strong_detection-for-different-epsilon_expand-at-start_pan-tilt-zoom}, we see similar results as for the independent learning with especially bad results for the maximal information coefficient and the distance correlation.

\begin{figure}	
	\centering
	\begin{subfigure}[t]{\textwidth}
		\centering
		\includegraphics[height=0.25\textheight]{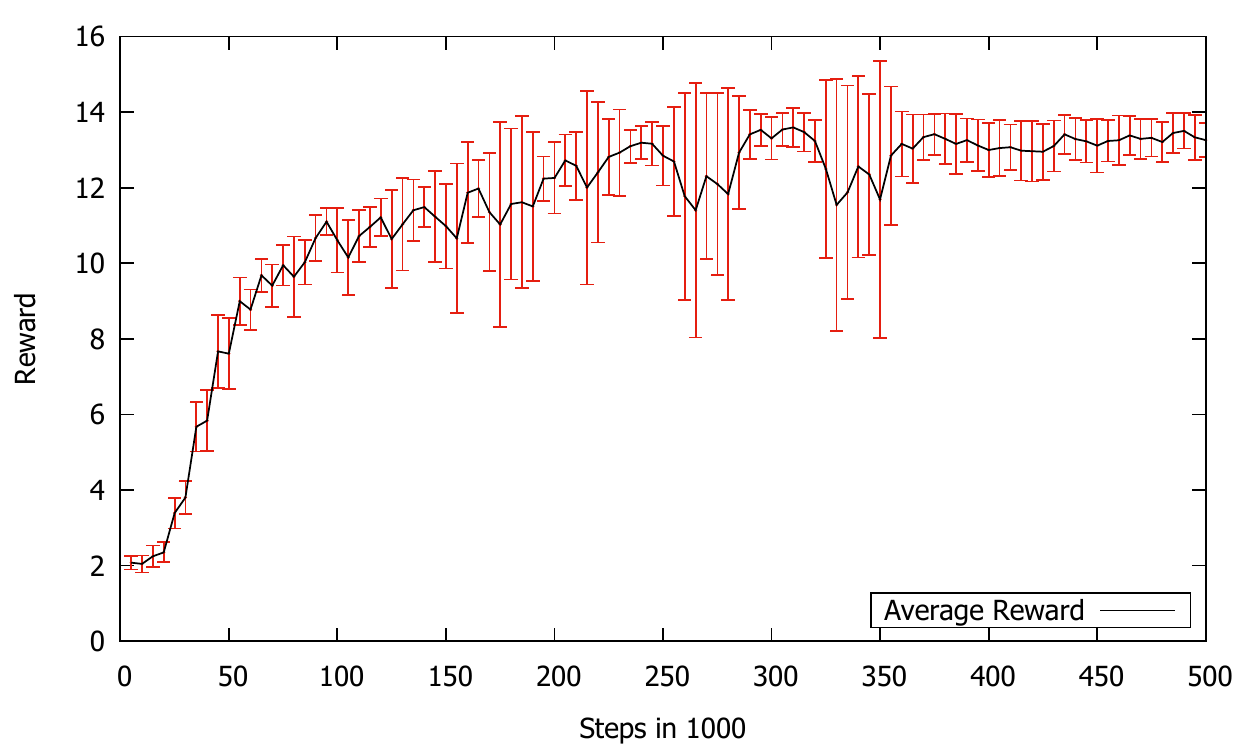}
		\caption{The results with single learners.}	
		\label{fig:300-detection-of-influences_further-aspects_adaption_detection-strong_reward_no-expand}	
	\end{subfigure}
	\\
	\begin{subfigure}[t]{\textwidth}
		\centering
		\includegraphics[height=0.25\textheight]{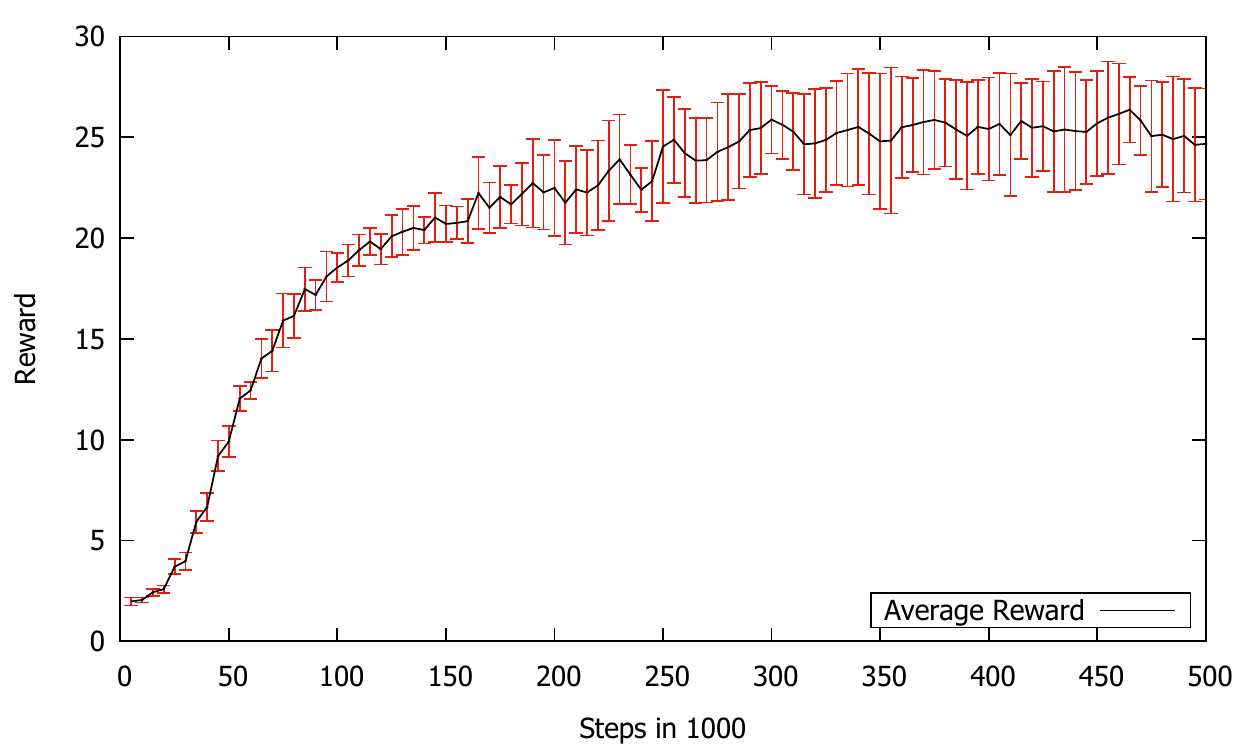}
		\caption{The results if the pan, tilt, and zoom configurations of Camera 2 are integrated in the state space of Camera 1 at the start of the simulation.}
		\label{fig:300-detection-of-influences_further-aspects_adaption_detection-strong_reward_expand-at-start}
	\end{subfigure}
	\\
	\begin{subfigure}[t]{\textwidth}
		\centering
		\includegraphics[height=0.25\textheight]{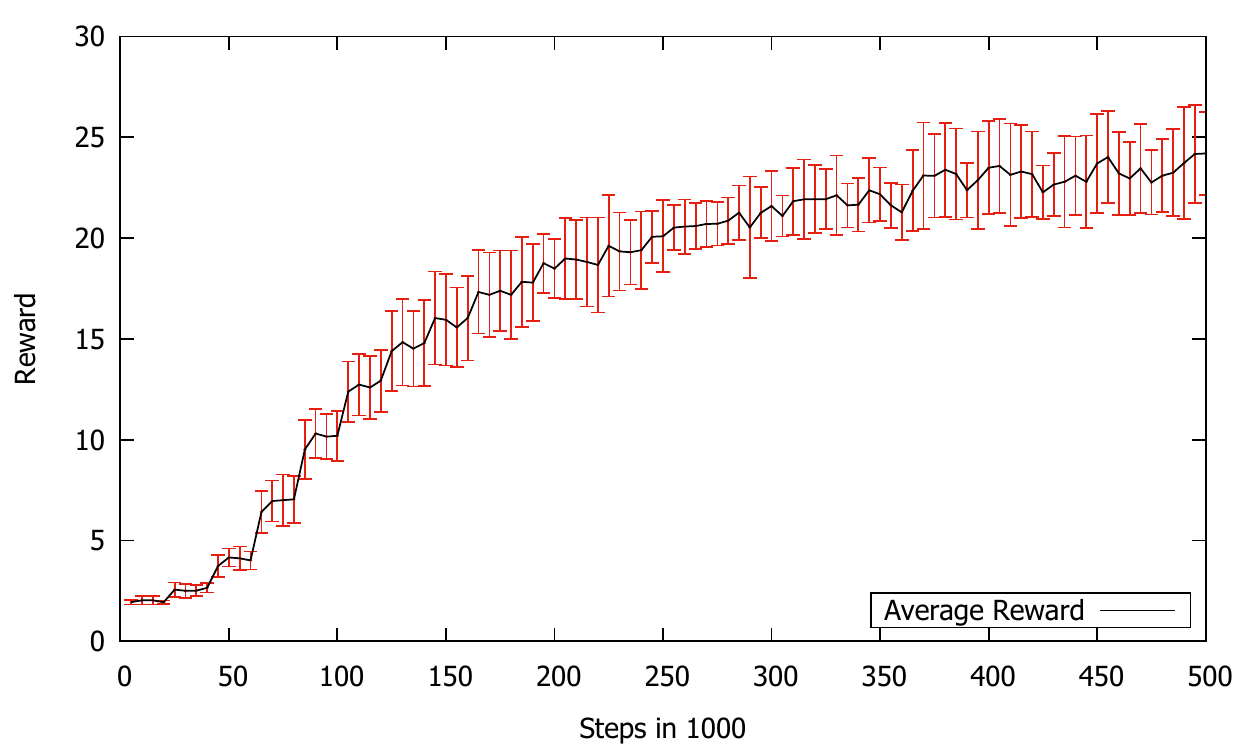}
		\caption{The results if the configurations are integrated in the state space of Camera 1 dynamically during runtime.}
		\label{fig:300-detection-of-influences_further-aspects_adaption_detection-strong_reward_dynamic-expand}
	\end{subfigure}
\caption[First part of comparison of the learning behavior on SCN 2.]{The figure shows the rewards received by the system if the Q-learning algorithm is applied to SCN 2 with single learners, with an integration at the start, and dynamic integration.}\label{fig:300-detection-of-influences_further-aspects_adaption_detection-strong_reward_no-expand-expand-at-start}
\end{figure}

As a last comparison on this scenario, we have a look at the rewards received with the different methods: the independent learners in Figure~\ref{fig:300-detection-of-influences_further-aspects_adaption_detection-strong_reward_no-expand}, the integration of the configuration of Camera 2 in the state of Camera 1 in Figure~\ref{fig:300-detection-of-influences_further-aspects_adaption_detection-strong_reward_expand-at-start}, and the dynamic integration of the configurations at runtime in Figure~\ref{fig:300-detection-of-influences_further-aspects_adaption_detection-strong_reward_dynamic-expand}. We can see that the independent learners are not able to reach an appropriate result since they lack a coordination mechanism. The integration at start can reach an optimal result of about $25$ after $250k$ steps. The dynamic expand can reach a similar result but learns slower. This is because the transfer mechanism cannot play out its advantages here since the Q-values differ for all states in this example.

\begin{figure}	
	\centering
	\includegraphics[width=0.5\textwidth]{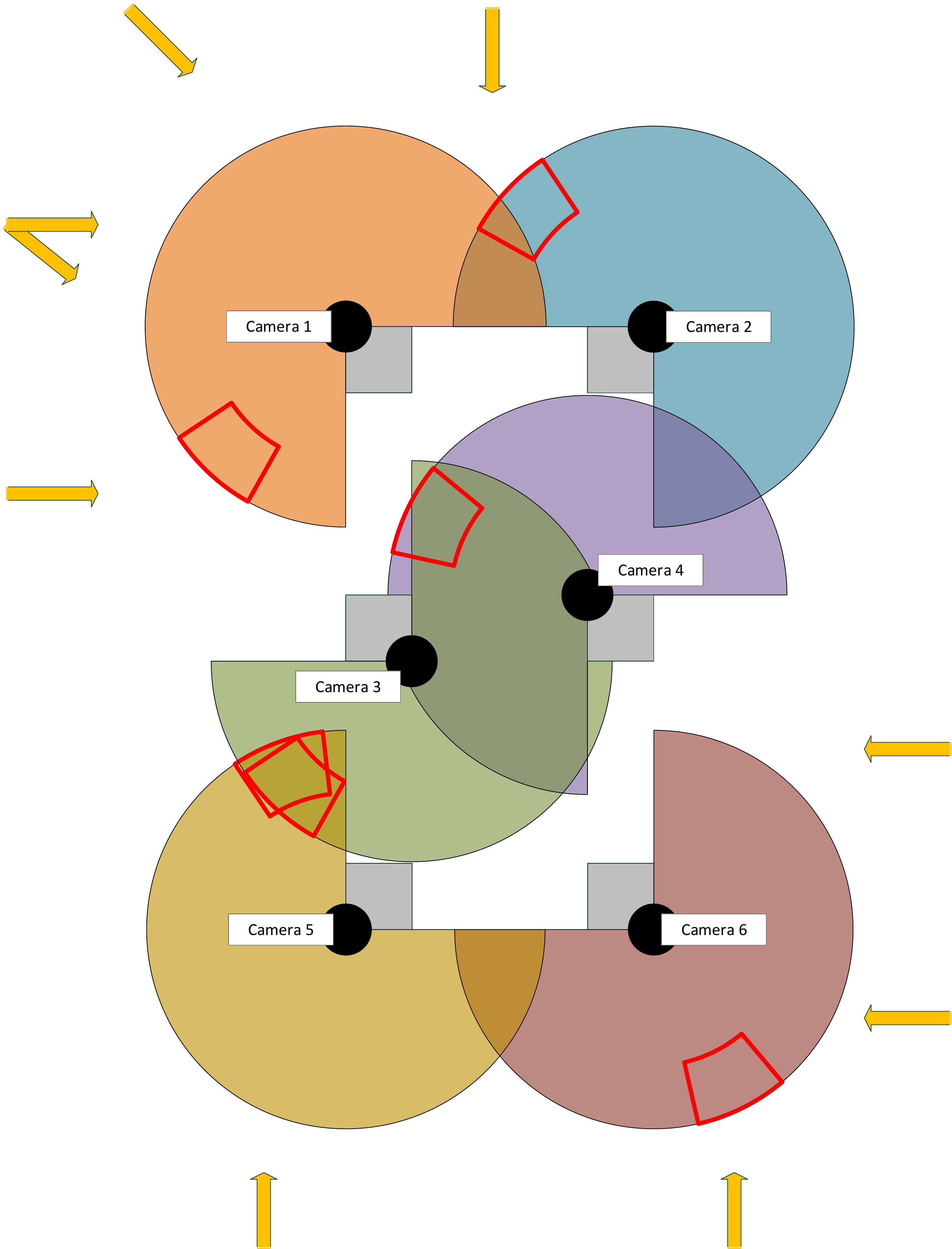}
	\caption[The smart camera scenario SCN 3.]{\textit{SCN 3}. A top-down view on a larger smart camera network with 6 cameras. The position of the cameras are depicted by the black dots. Each of them has a colored circle around it marking the area that can be potentially observed by it and a red shape that shows an exemplary observed area. The gray squares show the position of pillars limiting the potentially observable areas. The yellow arrows mark entry points of observation targets and there movement direction.}\label{fig:300-detection-of-influences_further-aspects_adaption_scenario3}
\end{figure}

In Figure~\ref{fig:300-detection-of-influences_further-aspects_adaption_scenario3}, we see scenario SCN 3, another scenario from the SC domain. The figure shows a top-down view on a SC network with six cameras depicted as black dots. Each of the cameras has a corresponding partial circle around it that visualizes the space potentially observable space of the cameras. The circle is only partial since the vision is blocked by pillars that are shown as gray squares. There are several streams of targets entering and crossing the scene from all sites marked by yellow arrows. We see that not each pair of cameras shares a common area both can observe. Therefore, only the pairs $(1,2)$, $(2,4)$, $(3,4)$, $(3,5)$, $(5,6)$ influence each other and have to consider the configuration of the other camera.

Again, we apply Q-learning to this scenario but with a small adaption: We do not use relative actions, e.g., increase or decrease, but absolute ones, i.e., each of the defined states can be accessed from every other state leading to $72$ actions. Since this does not require to learn sequences of action to reach one alignment from the previous the discount factor is set to $\gamma=0$.

\begin{figure}	
	\centering
	\begin{subfigure}[t]{\textwidth}
		\centering
		\includegraphics[height=0.25\textheight]{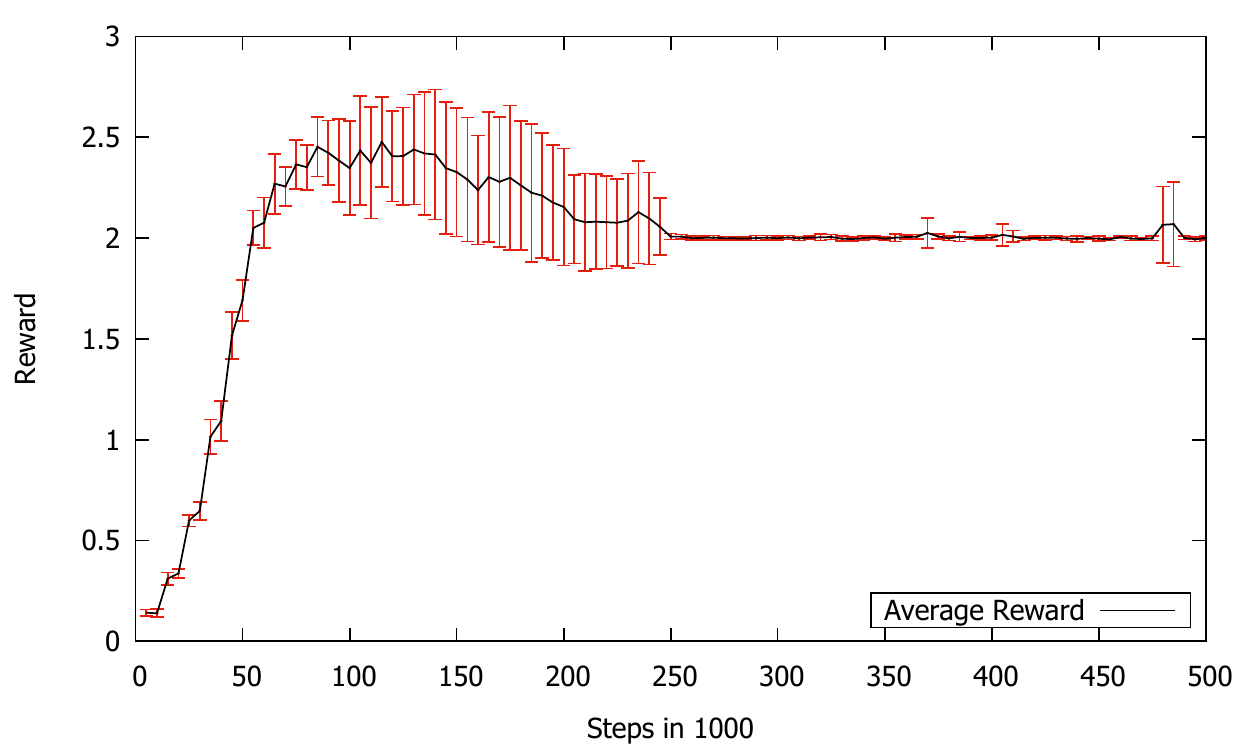}
		\caption{The cameras are controlled by single agent Q-learner.}	
		\label{fig:300-detection-of-influences_further-aspects_adaption_scenario3_reward_no-expand}	
	\end{subfigure}
	\\
	\begin{subfigure}[t]{\textwidth}
		\centering
		\includegraphics[height=0.25\textheight]{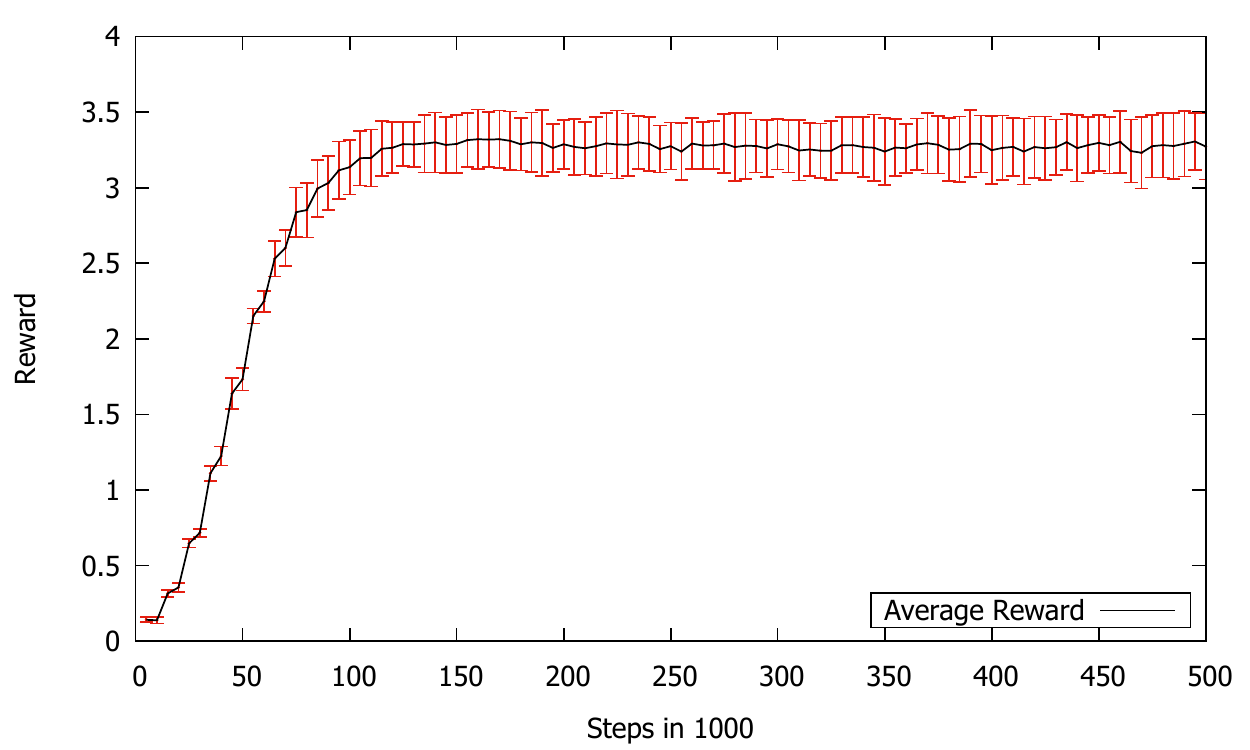}
		\caption{The results if $3k$ samples are used to determine the candidates for an integration.}	
		\label{fig:300-detection-of-influences_further-aspects_adaption_scenario3_reward_dynamic-expand_3k}	
	\end{subfigure}
	\\
	\begin{subfigure}[t]{\textwidth}
		\centering
		\includegraphics[height=0.25\textheight]{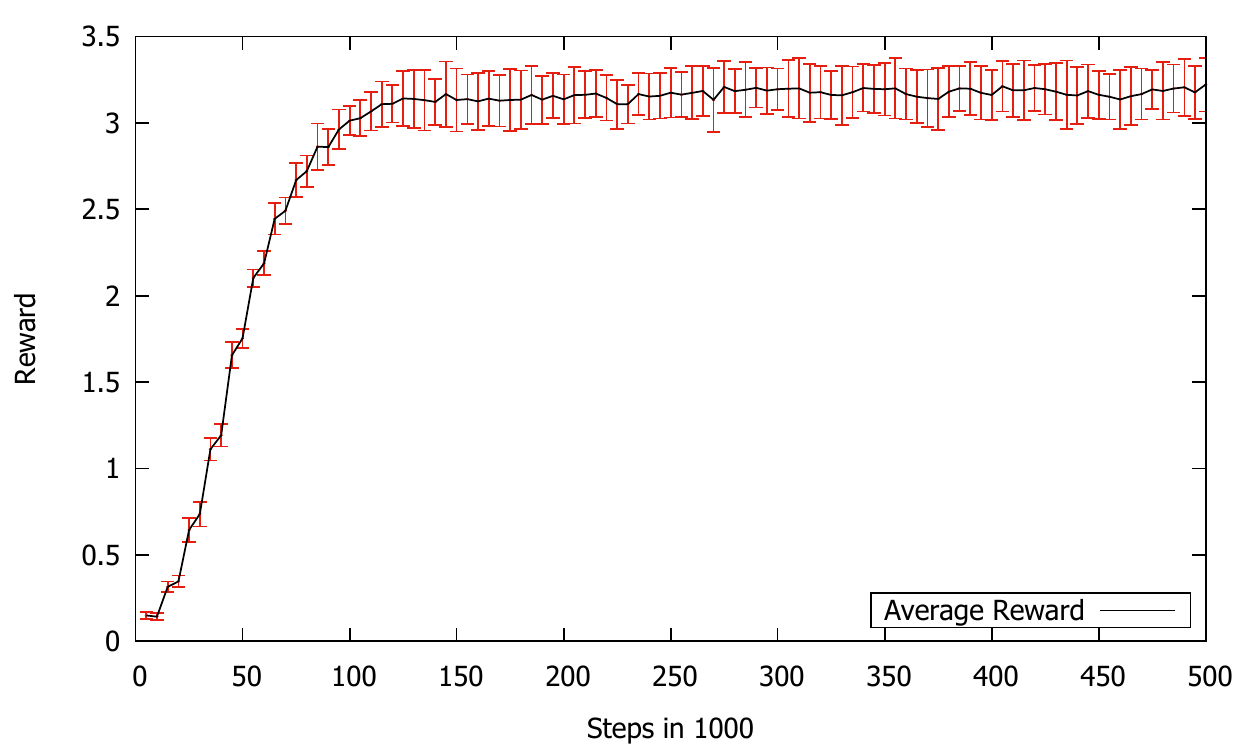}
		\caption{The results if $10k$ samples are used to determine the candidates for an integration.}
		\label{fig:300-detection-of-influences_further-aspects_adaption_scenario3_reward_dynamic-expand_10k}
	\end{subfigure}
\caption[Second part of comparison of the learning behavior on SCN 3.]{The figure shows the average reward over 10 runs in the scenario SCN 3. The cameras are each controlled by a Q-learning algorithm with a $\epsilon$-greedy action selection, where the $\epsilon$ decreases. It uses single learners or the integration of configurations at runtime.}\label{fig:300-detection-of-influences_further-aspects_adaption_scenario3_reward}
\end{figure}

The results are shown in Figure~\ref{fig:300-detection-of-influences_further-aspects_adaption_scenario3_reward}. The single-agent learning case in Figure~\ref{fig:300-detection-of-influences_further-aspects_adaption_scenario3_reward_no-expand} is clearly worse than the other variants only reaching a reward of $2$ on average since the cameras do not coordinate well. We see that the algorithm gets stuck in local optima when $\epsilon$ drops low at around $250k$ steps since single configurations are valued the highest due to the lack of knowledge of the influencing systems. The other two variants show the dynamic integration in the state space. In Figure~\ref{fig:300-detection-of-influences_further-aspects_adaption_scenario3_reward_dynamic-expand_3k}, we see how the system performs when $3,000$ samples are used for the detection; in Figure~\ref{fig:300-detection-of-influences_further-aspects_adaption_scenario3_reward_dynamic-expand_10k} $10,000$ samples are used. The graphs are nearly identical and both reach an optimal behavior that gives an average reward of above $3$.

\begin{figure}	
	\centering
	\resizebox{4cm}{!}{%
	\begin{tikzpicture}
  		\SetGraphUnit{5}
  		\Vertex{1}
  		\EA(1){2}
  		\SO(1){3}
  		\SO(3){5}
  		\SO(2){4}
  		\SO(4){6}
  		  		
  		\Edge[style={->}, label = PAN{,}TILT{,}ZOOM](1)(2)
  		\Edge[style={->}, label = PAN{,}TILT{,}ZOOM](3)(4)
  		\Edge[style={->}, label = PAN{,}TILT](5)(6)
  		\Edge[style={->}, label = PAN](5)(3)
  		\Edge[style={->}, label = PAN{,}TILT](2)(4)
  		
	\end{tikzpicture}
	}
	\caption{A snapshot of the structure of the learning network in an exemplary run on SCN 3 after $200000$ steps}\label{fig:300-detection-of-influences_further-aspects_adaption_scenario3_structure}
\end{figure}
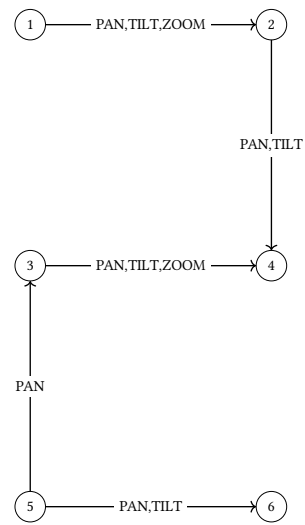

Finally, in Figure~\ref{fig:300-detection-of-influences_further-aspects_adaption_scenario3_structure}, we see the cooperation network that evolved after $200,000$ steps in one example run. Each arrow means that the configuration parts of the camera that is pointed at are integrated in the state space of the first camera. We see that the structure that formed reflects the structure of influences described before.

Concluding the results, we have seen how single learners compete with a-priori adapted systems and a stagewise integration of influencing configurations in the state space of an RL algorithm. We have seen that the single learner can not compete on the same level as influence respecting variants. The influence detecting approaches have shown similar results, which makes them useful options for the design of systems.

%
%
\section{Conclusion}\label{sec:conclusion}
Self-adaptive and self-organising systems typically act in a shared environment with other systems of the same kind. This results in the need to consider the current strategies followed by other systems within the own decision and learning process. We call such effects between actions and their effect on the achieved utility ``mutual influences''.

In this article, we presented a novel methodology to identify and consider these mutual influences by extending and refining prior work and investigating novel solutions to consider them in runtime self-adaptation. The detection is realized with dependency measures that have different characteristics and are consequently compared regarding their applicability in various settings. The adaption to the influences is realized using the example of Q-learning. The techniques are evaluated on two elementary use cases, the collaborative box manipulation and the two-man saw, and a smart camera application with multiple scenarios.

\textcolor{black}{
Finally, in the remaining paragraphs, we discuss the scalability of the approach and give an outlook on future work.
}

\subsection{Considerations about the Scalability of the Approach}
\textcolor{black}{
While the scalability is not the main focus of this work, we want to give an impression of this by considering and discussing two potential bottlenecks of the approach and afterwards give practical advice on how to handle it in situations where they might become an issue.
}

\textcolor{black}{
The first one is the computational overhead: Besides the calculations for the learning algorithm this leads to $n(n-1)k$ calculations of the dependency measures where n is the number of subsystems and k is the number of configuration components on each agent meaning the overhead will grow quadratic with the number of agents. However, this number will reduce vastly after the first components have been found influencing and are integrated in the learning algorithm since many possible paths are not considered any longer. For an experiment in the SC3 scenario with $n=6$ and $k=3$ this leads to a maximum of 90 calculations for each estimation of the influence. The experiments have been run sequentially on an Intel i5-2500K@3.3GHz and the approximate runtime for 150k steps (which is approximately the time until an optimal solution has been found) is about 1.9h including the time for the simulation itself and the learning algorithm. Given the remark before, it is clear that the average runtime of the influence estimation is much higher initially than in the end.\\
The second potential bottleneck is the distribution of the realizations of the configuration components. Depending on the system architecture, it is possible that each subsystem has to send its status to every other subsystem. Denoting the number of agents as $n$ and the number of components as $k$ this leads to $(n-1)k$ values send by each agent or $n(n-1)k$ values send in the whole system in each step. These values can have different size depending on the nature of the configuration space. Common value types are boolean, integer, and floating point numbers. Using the values of scenario SC3 with $n=6$ and $k=3$ this would result in 15 values send in each step by each agent.\\
While we believe that these requirements regarding the computational power and network capacity can be satisfied for a variety of systems that can benefit from our approach it is clear that some systems have limitations that might prevent the application of the influence detection. In this case, it is obviously possible to tweak the communication complexity by using algorithms that are more efficient given the systems architecture or to rely on less computational intensive dependency measures. Besides such implementation details, we want to discuss two basic ideas that can lead to a better scalability in general. These are the sequential analysis of systems and the reduction of network load. 
}
\textcolor{black}{
\paragraph{Sequential Analysis of Systems}
In the conducted experiments, each system measures the influence of each other system on itself. This can lead to a high system load and possibly to a backlog in the detection of influences, especially in large-scale systems.\\
To avoid such situations, it is possible to analyze each other system in a sequential order instead of simultaneously (possibly with multiple measures), i.e., a system creates an order and starts analyzing the influence from the first system in the sequence. As soon as it is done analyzing the first system it can switch to the next and so on. This order can be randomly or based on the physical or virtual distance, for instance. This reduces the computational peak load heavily.
}
\textcolor{black}{
\paragraph{Reduction of Network Load}
In the conducted experiments, we assumed that the influence origins from an arbitrary node in the overall system, i.e., each of the systems distributes its configuration via the communication network to the other systems. Depending on the connectivity of the systems, this can lead to a high network load affecting the communication necessary to operate the systems, e.g., the results in a wireless sensor network cannot be forwarded to the sink.\\
Similar to the reduction of computational resources, this can be avoided by analyzing the systems sequentially. Another effective way that can be used for some system types is to send the reward to the other systems instead of the configuration. This switches the places where the influences are calculated, i.e., instead of system $A$ receiving the configurations of system $B$ and calculating the influence of system $B$ on itself, it sends the rewards of itself to system $B$ that in turn calculates the influence. Since the configuration will most often create more traffic then just the reward this can reduce the computational overhead but this approach is limited to systems which are trusted and are willing to share their resources.
}

\subsection{Future Work}

In the following, two directions for future work are outlined. They aim mainly at further automating the detection process by an automatic combination of the dependency measures and to extend the applicability to systems that are not covered by the current system model.

\paragraph{Automated Weighting of Dependency Measures}
In this work, we used multiple dependency measures for the detection. This is due to the different detection rates for the various classes of systems. An interesting approach to extend the approach shown here is to combine the dependency measures to get a more reliable detection. This can be achieved by relying on simple techniques such as majority voting, but a useful approach could be to consider ensemble learning techniques~\cite{KrawczykMGSW2017}. Especially methods, where the combination of the learners is done by determining the weights for the individual learners on a case by case basis regarding the structure of the system is promising. The main challenge here lies in the engineering of a feedback signal or a training set that allows the combinator to find optimal weights for each situation.

\paragraph{Detecting Influences in Systems without a Local Reward Signal}
In this article, the system model relies on a local reward that is a feedback signal to the subsystems regarding their decisions. If there is no obvious choice for such a reward signal, it can be useful to have an influence detection mechanism that is able to work without such. Therefore, an interesting approach for future work is to change the system model to a type where only the configuration and a current state of the subsystems is necessary. The dependency can then be measured between one system's configuration and another system's situation. However, this introduces the problem that the situation is very likely not a one dimensional variable but has several dimensions, e.g., the position is composed of an $x$- and $y$-coordinate. Therefore, it will be necessary to use dimension reduction techniques on the situation of a system or to measure the dependency between two groups of random variables, e.g., with a canonical correlation analysis~\cite{HaerdleS2007}.

Besides these two concrete directions it is promising to apply the influence detection to further real-world systems, especially with different learning algorithms and influence structures.

\begin{acks}
	
We gratefully acknowledge the financial support of the German Research Foundation (DFG) in the context of the project CYPHOC (HA 5480/3-1,SI 674/9-1).

Furthermore, we would like to thank our project partners Bernhard Sick, Christian Gruhl, Henner Heck (Uni Kassel, IES group) and Arno Wacker (Universit\"at der Bundeswehr M\"unchen) for the fruitful discussions in the project.

\end{acks}


\bibliographystyle{ieeetr}
\bibliography{diss.bib}

\begin{thebibliography}{10}

\bibitem{KephartC2003a}
J.~O. Kephart and D.~M. Chess, ``The vision of autonomic computing,'' vol.~36,
  no.~1, pp.~41--50.

\bibitem{Mueller-SchloerT2018}
C.~M{\"u}ller-Schloer and S.~Tomforde, {\em Organic Computing -- Technical
  Systems for Survival in the Real World}.
\newblock Springer International Publishing, 2018.

\bibitem{Tennenhouse2000}
D.~Tennenhouse, ``Proactive computing,'' vol.~43, no.~5, pp.~43--50.

\bibitem{Wooldridge2009}
M.~Wooldridge, {\em An introduction to multiagent systems}.
\newblock John Wiley \& Sons.

\bibitem{KernbachST2011}
S.~Kernbach, T.~Schmickl, and J.~Timmis, ``Collective adaptive systems:
  Challenges beyond evolvability,'' vol.~abs/1108.5643.

\bibitem{Weyns2013}
D.~Weyns, B.~Schmerl, V.~Grassi, S.~Malek, R.~Mirandola, C.~Prehofer,
  J.~Wuttke, J.~Andersson, H.~Giese, and K.~M. G{\"o}schka, ``On patterns for
  decentralized control in self-adaptive systems,'' in {\em Software
  Engineering for Self-Adaptive Systems II}, pp.~76--107, Springer, 2013.

\bibitem{TomfordeM2014}
S.~Tomforde and C.~M{\"{u}}ller{-}Schloer, ``Incremental design of adaptive
  systems,'' vol.~6, no.~2, pp.~179--198.

\bibitem{Krupitzer2015}
C.~Krupitzer, F.~M. Roth, S.~VanSyckel, G.~Schiele, and C.~Becker, ``A survey
  on engineering approaches for self-adaptive systems,'' {\em Pervasive and
  Mobile Computing}, vol.~17, pp.~184--206, 2015.

\bibitem{Maier1998}
M.~W. Maier, ``Architecting principles for systems-of-systems,'' {\em Systems
  Engineering}, vol.~1, no.~4, pp.~267--284, 1998.

\bibitem{TomfordeHSRSWS2014}
S.~Tomforde, J.~H{\"{a}}hner, H.~Seebach, W.~Reif, B.~Sick, A.~Wacker, and
  I.~Scholtes, ``{Engineering and Mastering Interwoven Systems},'' in {\em
  {ARCS} 2014 - 27th International Conference on Architecture of Computing
  Systems, Workshop Proceedings, February 25-28, 2014, Luebeck, Germany,
  University of Luebeck, Institute of Computer Engineering}, pp.~1--8.

\bibitem{TomfordeRBW2016}
S.~Tomforde, S.~Rudolph, K.~L. Bellman, and R.~P. W{\"{u}}rtz, ``An organic
  computing perspective on self-improving system interweaving at runtime,'' in
  {\em 2016 {IEEE} International Conference on Autonomic Computing, {ICAC}
  2016, Wuerzburg, Germany, July 17-22, 2016}, pp.~276--284.

\bibitem{RudolphETH2014}
S.~Rudolph, S.~Edenhofer, S.~Tomforde, and J.~H{\"{a}}hner, ``Reinforcement
  learning for coverage optimization through {PTZ} camera alignment in highly
  dynamic environments,'' in {\em Proceedings of the International Conference
  on Distributed Smart Cameras, {ICDSC} '14, Venezia Mestre, Italy, November
  4-7, 2014}, pp.~19:1--19:6.

\bibitem{Piciarelli2016}
C.~Piciarelli, L.~Esterle, A.~Khan, B.~Rinner, and G.~L. Foresti, ``Dynamic
  reconfiguration in camera networks: A short survey,'' {\em IEEE Transactions
  on Circuits and Systems for Video Technology}, vol.~26, no.~5, pp.~965--977,
  2016.

\bibitem{Bellman2018}
K.~Bellman, J.~Botev, A.~Diaconescu, L.~Esterle, C.~Gruhl, C.~Landauer,
  P.~Lewis, A.~Stein, S.~Tomforde, and R.~Würtz, ``Self-improving system
  integration--status and challenges after five years of sissy,'' 2018.

\bibitem{RudolphHTH2016}
S.~Rudolph, R.~Hihn, S.~Tomforde, and J.~H{\"{a}}hner, ``Comparison of
  dependency measures for the detection of mutual influences in organic
  computing systems,'' in {\em Architecture of Computing Systems - {ARCS} 2016
  - 29th International Conference, Nuremberg, Germany, April 4-7, 2016,
  Proceedings}, pp.~334--347.

\bibitem{RudolphTH2016}
S.~Rudolph, S.~Tomforde, and J.~Hähner, ``A mutual influence-based learning
  algorithm,'' in {\em Proceedings of the 8th International Conference on
  Agents and Artificial Intelligence {(ICAART} 2016), Volume 1, Rome, Italy,
  February 24-26, 2016.}, pp.~181--189.

\bibitem{RudolphTSH2015}
S.~Rudolph, S.~Tomforde, B.~Sick, and J.~H{\"{a}}hner, ``A mutual influence
  detection algorithm for systems with local performance measurement,'' in {\em
  2015 {IEEE} 9th International Conference on Self-Adaptive and Self-Organizing
  Systems, Cambridge, MA, USA, September 21-25, 2015}, pp.~144--149.

\bibitem{THH10-a}
S.~Tomforde, B.~Hurling, and J.~H{\"a}hner, ``{Dynamic control of mobile ad-hoc
  networks - Network protocol parameter adaptation using Organic Network
  Control},'' in {\em {Proc. of the 7th Int. Conf. on Informatics in Control,
  Automation, and Robotics (ICINCO'10), held in Funchal, Portugal (June 15 -
  18, 2010)}}, (Setubal), pp.~28--35, INSTICC, 2010.

\bibitem{SuttonB1998}
R.~S. Sutton and A.~G. Barto, {\em Introduction to Reinforcement Learning}.
\newblock MIT Press, 1st~ed.

\bibitem{Lamsweerde2001}
A.~van Lamsweerde, ``Goal-oriented requirements engineering: a guided tour,''
  in {\em Proceedings Fifth IEEE International Symposium on Requirements
  Engineering}, pp.~249--262.

\bibitem{Lamsweerde2009}
A.~van Lamsweerde, {\em Requirements Engineering: From System Goals to UML
  Models to Software Specifications}.
\newblock Wiley Publishing, 1st~ed.

\bibitem{FredericksDC2014}
E.~M. Fredericks, B.~DeVries, and B.~H.~C. Cheng, ``Towards run-time adaptation
  of test cases for self-adaptive systems in the face of uncertainty,'' in {\em
  Proceedings of the 9th International Symposium on Software Engineering for
  Adaptive and Self-Managing Systems}, SEAMS 2014, (New York, NY, USA),
  pp.~17--26, ACM.

\bibitem{TH11-a}
S.~Tomforde and J.~H\"ahner, {\em {Biologically Inspired Networking and
  Sensing: Algorithms and Architectures}}, ch.~{Organic Network Control --
  Turning Standard Protocols Into Evolving Systems}, pp.~11 -- 35.
\newblock IGI, 2011.

\bibitem{BusoniuBD2008}
L.~Busoniu, R.~Babuska, and B.~De~Schutter, ``A comprehensive survey of
  multiagent reinforcement learning,'' vol.~38, no.~2, pp.~156--172.

\bibitem{WieringO2014}
M.~Wiering and M.~van Otterlo, {\em Reinforcement Learning: State-of-the-Art}.
\newblock Springer Publishing Company, Incorporated.

\bibitem{WatkinsD1992}
C.~J. C.~H. Watkins and P.~Dayan, ``Technical note q-learning.,'' vol.~8,
  pp.~279--292.

\bibitem{RudolphTSHWH2015}
S.~Rudolph, S.~Tomforde, B.~Sick, H.~Heck, A.~Wacker, and J.~Hähner, ``An
  online influence detection algorithm for organic computing systems,'' in {\em
  ARCS 2015 - The 28th International Conference on Architecture of Computing
  Systems. Proceedings}, pp.~1--8.

\bibitem{KeilG2003}
D.~Keil and D.~Q. Goldin, ``Modeling indirect interaction in open computational
  systems,'' in {\em 12th {IEEE} International Workshops on Enabling
  Technologies {(WETICE} 2003), Infrastructure for Collaborative Enterprises,
  9-11 June 2003, Linz, Austria}, pp.~371--376.

\bibitem{LogieHW2008}
R.~Logie, J.~G. Hall, and K.~G. Waugh, ``Towards mining for influence in a
  multi agent environment.,'' in {\em IADIS European Conf. Data Mining}
  (A.~Abraham, ed.), pp.~97--101, IADIS.

\bibitem{LogieHW2010}
R.~Logie, J.~G. Hall, and K.~G. Waugh, ``Investigating agent influence and
  nested other-agent behaviour,'' {\em International Journal on Advances in
  Intelligent Systems Volume 2, Number 4, 2009}, 2010.

\bibitem{Broersen2010}
J.~M. Broersen, ``{CTL.STIT:} enhancing {ATL} to express important multi-agent
  system verification properties,'' in {\em 9th International Conference on
  Autonomous Agents and Multiagent Systems {(AAMAS} 2010), Toronto, Canada, May
  10-14, 2010, Volume 1-3}, pp.~683--690.

\bibitem{StoneV2000}
P.~Stone and M.~Veloso, ``Multiagent systems: A survey from a machine learning
  perspective,'' vol.~8, no.~3, pp.~345--383.

\bibitem{KokSV2003}
J.~R. Kok, M.~T.~J. Spaan, and N.~Vlassis, ``Multi-robot decision making using
  coordination graphs,'' in {\em Proceedings of the International Conference on
  Advanced Robotics (ICAR)} (A.~T. de~Almeida and U.~Nunes, eds.),
  pp.~1124--1129.

\bibitem{KokHBV2005}
J.~R. Kok, P.~J. 't~Hoen, B.~Bakker, and N.~Vlassis, ``Utile coordination:
  learning interdependencies among cooperative agents,'' in {\em Proceedings of
  the IEEE Symposium on Computational Intelligence and Games (CIG)},
  pp.~29--36.

\bibitem{DeHauwereVN2009}
Y.-M. De~Hauwere, P.~Vrancx, and A.~Now{\'e}, ``Learning what to observe in
  multi-agent systems,'' in {\em Proceedings of the 20th Belgian-Netherlands
  Conference on Artificial Intelligence}, pp.~83--90, 2009.

\bibitem{DeHauwereVN2010}
Y.-M. De~Hauwere, P.~Vrancx, and A.~Nowé, ``Learning multi-agent state space
  representations,'' in {\em Proceedings of the 9th International Conference on
  Autonomous Agents and Multiagent Systems: Volume 1}, pp.~715--722,
  International Foundation for Autonomous Agents and Multiagent Systems.

\bibitem{DeHauwereVN2011}
Y.-M. De~Hauwere, P.~Vrancx, and A.~Now{\'e}, ``Solving delayed coordination
  problems in mas,'' in {\em The 10th International Conference on Autonomous
  Agents and Multiagent Systems - Volume 3}, AAMAS '11, (Richland, SC),
  pp.~1115--1116, International Foundation for Autonomous Agents and Multiagent
  Systems.

\bibitem{Lanctot2017}
M.~Lanctot, V.~Zambaldi, A.~Gruslys, A.~Lazaridou, K.~Tuyls, J.~P{\'e}rolat,
  D.~Silver, and T.~Graepel, ``A unified game-theoretic approach to multiagent
  reinforcement learning,'' in {\em Advances in Neural Information Processing
  Systems}, pp.~4190--4203, 2017.

\bibitem{PengLD2005}
H.~Peng, F.~Long, and C.~Ding, ``Feature selection based on mutual information:
  Criteria of max-dependency, max-relevance, and min-redundancy,'' vol.~27,
  no.~8, pp.~1226--1238.

\bibitem{FarahmandGSM2008}
A.~M. Farahmand, M.~Ghavamzadeh, C.~Szepesv{\'{a}}ri, and S.~Mannor,
  ``Regularized policy iteration,'' in {\em Advances in Neural Information
  Processing Systems 21, Proceedings of the Twenty-Second Annual Conference on
  Neural Information Processing Systems, Vancouver, British Columbia, Canada,
  December 8-11, 2008}, pp.~441--448.

\bibitem{KolterN2009}
J.~Z. Kolter and A.~Y. Ng, ``Regularization and feature selection in
  least-squares temporal difference learning,'' in {\em Proceedings of the 26th
  Annual International Conference on Machine Learning}, ICML '09, (New York,
  NY, USA), pp.~521--528, ACM, 2009.

\bibitem{LiuLW2015}
D.-R. Liu, H.-L. Li, and D.~Wang, ``Feature selection and feature learning for
  high-dimensional batch reinforcement learning: A survey,'' vol.~12, no.~3,
  pp.~229--242.

\bibitem{Boes2017}
J.~Boes and F.~Migeon, ``Self-organizing multi-agent systems for the control of
  complex systems,'' {\em Journal of Systems and Software}, vol.~134,
  pp.~12--28, 2017.

\bibitem{RudolphHTH2017}
S.~Rudolph, R.~Hihn, S.~Tomforde, and J.~Hähner, ``Towards discovering delayed
  mutual influences in organic computing systems,'' in {\em ARCS 2017; 30th
  GI/ITG International Conference on Architecture of Computing Systems},
  pp.~39--46.

\bibitem{Pearson1895}
K.~Pearson, ``{Note on regression and inheritance in the case of two
  parents},'' vol.~58, no.~347-352, pp.~240--242.

\bibitem{Kendall1938}
M.~G. Kendall, ``A new measure of rank correlation,'' {\em Biometrika},
  vol.~30, no.~1/2, pp.~81--93, 1938.

\bibitem{SzekelyRB2007}
G.~J. Sz\'{e}kely, M.~L. Rizzo, and N.~K. Bakirov, ``{Measuring and testing
  dependence by correlation of distances},'' vol.~35, no.~6, pp.~2769--2794.

\bibitem{ShannonW1949}
C.~Shannon and W.~Weaver, {\em The Mathematical Theory of Communication}.
\newblock University of Illinois Press.

\bibitem{KraskovSG2004}
A.~Kraskov, H.~St\"ogbauer, and P.~Grassberger, ``Estimating mutual
  information,'' vol.~69, p.~066138.

\bibitem{ReshefRFGMTLMS2011}
D.~N. Reshef, Y.~A. Reshef, H.~K. Finucane, S.~R. Grossman, G.~McVean, P.~J.
  Turnbaugh, E.~S. Lander, M.~Mitzenmacher, and P.~C. Sabeti, ``Detecting novel
  associations in large data sets,'' vol.~334, no.~6062, pp.~1518--1524.

\bibitem{LukeCPSB2005}
S.~Luke, C.~Cioffi-Revilla, L.~Panait, K.~Sullivan, and G.~Balan, ``Mason: A
  multiagent simulation environment,'' {\em Simulation: Transactions of the
  society for Modeling and Simulation International}, vol.~82, no.~7,
  pp.~517--527, 2005.

\bibitem{TaylorS2009}
M.~E. Taylor and P.~Stone, ``Transfer learning for reinforcement learning
  domains: A survey,'' {\em Journal of Machine Learning Research}, vol.~10,
  no.~Jul, pp.~1633--1685, 2009.

\bibitem{Wilson1995}
S.~W. Wilson, ``{Classifier Fitness Based on Accuracy},'' vol.~3, no.~2,
  pp.~149--175.

\bibitem{Wilson2000}
S.~W. Wilson, ``Get real! xcs with continuous-valued inputs,'' in {\em Learning
  Classifier Systems} (P.~L. Lanzi, W.~Stolzmann, and S.~W. Wilson, eds.),
  (Berlin, Heidelberg), pp.~209--219, Springer Berlin Heidelberg, 2000.

\bibitem{LiY2016}
X.~Li and G.~Yang, ``Transferable xcs,'' in {\em Proceedings of the Genetic and
  Evolutionary Computation Conference 2016}, GECCO '16, (New York, NY, USA),
  pp.~453--460, ACM, 2016.

\bibitem{KrawczykMGSW2017}
B.~Krawczyk, L.~L. Minku, J.~Gama, J.~Stefanowski, and M.~Wo{\'z}niak,
  ``Ensemble learning for data stream analysis: A survey,'' {\em Information
  Fusion}, vol.~37, pp.~132--156, 2017.

\bibitem{HaerdleS2007}
W.~H{\"a}rdle and L.~Simar, {\em Applied multivariate statistical analysis},
  vol.~22007.
\newblock Springer, 2007.

\end{thebibliography}

\end{document}